\documentclass[preprint, aps, prd]{revtex4-1}

\def\mc{\mathcal}
\def\ul{\underline}

\usepackage[dvips]{graphicx}
\usepackage{amssymb}
\usepackage{amssymb,amsmath}
\usepackage{graphicx}
\usepackage{dsfont}
\usepackage{caption}
\usepackage{subcaption}
\usepackage{mathtools}
\usepackage{verbatim}
\usepackage{graphicx}
\usepackage{multirow}
\usepackage[outline]{contour}
\usepackage{xcolor,colortbl}
\makeatother
\begin{document}

\title{Supersymmetric domain walls in maximal 6D gauged supergravity II}

\author{Parinya Karndumri} \author{Patharadanai Nuchino} \email[REVTeX Support:
]{parinya.ka@hotmail.com and danai.nuchino@hotmail.com} 
\affiliation{String Theory and
Supergravity Group, Department of Physics, Faculty of Science,
Chulalongkorn University, 254 Phayathai Road, Pathumwan, Bangkok
10330, Thailand}

\date{\today}
\begin{abstract}
We continue the study of supersymmetric domain wall solutions in six-dimensional maximal gauged supergravity. We first give a classification of viable gauge groups with the embedding tensor in $\mathbf{5}^{+7}$, $\bar{\mathbf{5}}^{+3}$, $\mathbf{10}^{-1}$, $\mathbf{24}^{-5}$, and $\overline{\mathbf{45}}^{+3}$ representations of the off-shell symmetry $GL(5)\subset SO(5,5)$. We determine an explicit form of the embedding tensor for gauge groups arising from each representation together with some examples of possible combinations among them. All of the resulting gauge groups are of a non-semisimple type with abelian factors and translational groups of different dimensions. We find $\frac{1}{2}$- and $\frac{1}{4}$-supersymmetric domain walls with $SO(2)$ symmetry in $SO(2)\ltimes \mathbb{R}^8$ and $SO(2)\ltimes \mathbb{R}^6$ gauge groups from the embedding tensor in $\mathbf{24}^{-5}$ representation and in $CSO(2,0,2)\ltimes \mathbb{R}^4$, $CSO(2,0,2)\ltimes \mathbb{R}^2$, and $CSO(2,0,1)\ltimes \mathbb{R}^4$ gauge groups with the embedding tensor in $\overline{\mathbf{45}}^{+3}$ representations. These gauge groups are parametrized by a traceless matrix and electrically and magnetically embedded in $SO(5,5)$ global symmetry, respectively.
\end{abstract}
\maketitle

\section{Introduction}
Domain wall solutions in gauged supergravities play an important role in high-energy physics. In the AdS/CFT correspondence \cite{maldacena,Gubser_AdS_CFT,Witten_AdS_CFT} and a generalization to non-conformal field theories called the DW/QFT correspondence \cite{DW_QFT1,DW_QFT2,DW_QFT3,DW_QFT4}, these solutions give holographic descriptions of RG flows in the dual conformal and non-conformal field theories at strong coupling, see for example \cite{GPPZ,FGPP,KW_5Dflow,N2_IIB_flow,AdS_RG_flow,N1_SYM_fixed_point}. Domain walls are also useful in studying cosmology \cite{DW_cosmology1,DW_cosmology2,DW_cosmology3,DW_cosmology4}, see also \cite{DW_inflation} for a recent result. Most of the solutions are supersymmetric and have been found within the framework of gauged supergravities in various dimensions, see \cite{Eric_DW_10D,9D_DW_Cowdall,9D_DW,8D_DW1,8D_DW2,7D_DW,6D_DW_I,Pope_DW_massive,DW_algebraic_curve,
Eric_DW_maximal_SUGRA,Pope_symmetric_poten,DW_Hull,DW_Cvetic,4D_DW,3DN4_DW,3DN10_DW,2D_DW1,2D_DW2,Eric_SUSY_DW} for an incomplete list. 
\\
\indent In this paper, we are interested in supersymmetric domain walls from the maximal $N=(2,2)$ gauged supergravity in six dimensions constructed in \cite{6D_Max_Gauging}, see \cite{6D_SO(5)} for an earlier result. The result of \cite{6D_Max_Gauging} describes the most general gaugings of $N=(2,2)$ supergravity in six dimensions using the embedding tensor formalism. The embedding tensor lives in representation $\mathbf{144}_c$ of the global symmetry $SO(5,5)$ and determines a viable gauge group $G_0\subset SO(5,5)$. In general, there are a large number of possible gauge groups. In this work, we will consider only gauge groups classified under $GL(5)\subset SO(5,5)$ which is an off-shell symmetry of the $N=(2,2)$ supergravity Lagrangian in a particular symplectic frame. 
\\
\indent We will study various possible gaugings and explicitly construct the corresponding embedding tensors for the resulting gauge groups. We will also look for supersymmetric domain wall solutions. According to the DW/QFT correspondence, these solutions are dual to maximally supersymmetric Yang-Mills theory (SYM) in five dimensions which recently plays an important role in studying dynamics of (conformal) field theories in higher and lower dimensions via a number of dualities, see for example \cite{Douglas_5D_MSYM,5D_MSYM_1,5D_MSYM_2,5D_MSYM_3,5D_MSYM_4,5D_MSYM_5}. The five-dimensional SYM could be used to define $N=(2,0)$ superconformal field theory (SCFT) in six dimensions compactified on $S^1$. Since the latter is well-known to describe dynamics of strongly coupled theory on M5-branes, we expect that supersymmetric domain walls of the maximal gauged supergravity in six dimensions could be useful in studying various aspects of the maximal SYM in five dimensions as well as six-dimensional SCFT and physics of M5-branes at strong coupling. In addition, five-dimensional maximal SYM and compactifications on $S^1$ and $S^2$ can lead to some insights to S-duality of twisted gauge theories in four dimensions and monopoles in Aharony-Bergman-Jafferis-Maldacena (ABJM) theory. Therefore, the resulting domain walls could also be useful in this context as well. However, it should be pointed out that all gaugings classified here currently have no known higher dimensional origins. 
\\
\indent Since $N=4$ superconformal symmetry in five dimensions does not exist \cite{Nahm_res}, see also a recent classification of maximally supersymmetric $AdS$ vacua given in \cite{Max_SUSY_AdS}, there is no AdS$_6$/CFT$_5$ duality with $32$ supercharges. Accordingly, supersymmetric domain walls in $N=(2,2)$ gauged supergravity are expected to play a prominent role in holographic study in this case. A domain wall solution with $SO(5)$ symmetry in $SO(5)$ gauging has been found in \cite{6D_SO(5)} with the holographic interpretation given in \cite{Bobev_spherical_brane1} and \cite{Bobev_spherical_brane2}. Moreover, a large number of domain wall solutions has been given recently in \cite{6D_DW_I} with various gauge groups obtained from the embedding tensor in $\mathbf{15}^{-1}$ and $\overline{\mathbf{40}}^{-1}$ representations of $GL(5)$. We will extend this investigation by considering the embedding tensor in other representations of $GL(5)$. These are given by $\mathbf{5}^{+7}$, $\bar{\mathbf{5}}^{+3}$, $\mathbf{10}^{-1}$, $\mathbf{24}^{-5}$, and $\overline{\mathbf{45}}^{+3}$ representations. It turns out that most of the resulting gauge groups are of non-semisimple type without any compact subgroup. Due to the complexity of working with the full $25$-dimensional scalar manifold $SO(5,5)/SO(5)\times SO(5)$, we mainly follow the method introduced in \cite{New_Extrema} to reduce the number of scalar fields to make the analysis more traceable. However, this approach requires the existence of a non-trivial symmetry $H_0\subset G_0$. In many of the gaugings classified here, the residual $H_0$ is absent, so there are too many scalars to take into account. Accordingly, we will give domain wall solutions only for gauge groups with at least $SO(2)$ subgroup. As in \cite{6D_DW_I}, there exist both $\frac{1}{2}$- and $\frac{1}{4}$-supersymmetric domain wall solutions in accord with the general classification of supersymmetric domain walls given in \cite{Eric_SUSY_DW}.
\\
\indent The paper is organized as follows. In section \ref{6DN=(2,2)gSUGRA}, we briefly review the construction of six-dimensional maximal gauged supergravity in the embedding tensor formalism. Gaugings in $\mathbf{5}^{+7}$, $\bar{\mathbf{5}}^{+3}$, $\mathbf{10}^{-1}$, $\mathbf{24}^{-5}$, and $\overline{\mathbf{45}}^{+3}$ representations of $GL(5)$ are classified in section \ref{gauging}. In section \ref{DW_sol}, we explicitly construct a number of supersymmetric domain wall solutions. Conclusions and discussions are given in section \ref{Discuss}, and useful formulae are given in the appendix.

\section{$N=(2,2)$ gauged supergravity in six dimensions}\label{6DN=(2,2)gSUGRA}
We first give a review of six-dimensional $N=(2,2)$ gauged supergravity in the embedding tensor formalism constructed in \cite{6D_Max_Gauging}. We will mainly collect relevant formulae for constructing the embedding tensor in order to classify various gauge groups and find supersymmetric domain wall solutions. For more details, the reader is referred to the original construction in \cite{6D_Max_Gauging}.

There is only one supermultiplet in $N=(2,2)$ supersymmetry in six dimensions, the graviton supermultiplet, with the following field content
\begin{equation}\label{6DSUGRAmultiplet}
\left(e^{\hat{\mu}}_\mu, B_{\mu\nu m}, A^{A}_\mu, {V_A}^{\alpha\dot{\alpha}}, \psi_{+\mu\alpha}, \psi_{-\mu\dot{\alpha}}, \chi_{+a\dot{\alpha}}, \chi_{-\dot{a}\alpha}\right).
\end{equation}
We note here all the conventions used throughout the paper. These mostly follow those used in \cite{6D_Max_Gauging}. Curved and flat space-time indices are respectively denoted by $\mu,\nu,\ldots=0,1,\ldots,5$ and $\hat{\mu},\hat{\nu},\ldots=0,1,\ldots,5$. Lower and upper $m,n,\ldots=1,\ldots,5$ indices label fundamental and anti-fundamental representations of $GL(5)\subset SO(5,5)$, respectively. Indices $A,B,\ldots =1,\ldots,16$ describe Majorana-Weyl spinors of the $SO(5,5)$ duality symmetry. We also note that the electric two-form potentials $B_{\mu\nu m}$, appearing in the ungauged Lagrangian, transform as $\mathbf{5}$ under $GL(5)$ while the vector fields $A^{A}_\mu$ transform as $\mathbf{16}_c$ under $SO(5,5)$. Together with the magnetic duals ${B_{\mu\nu}}^m$ transforming in $\overline{\mathbf{5}}$ representation of $GL(5)$, the electric two-forms $B_{\mu\nu m}$ transform in a vector representation $\mathbf{10}$ of the full global symmetry group $SO(5,5)$ denoted by $B_{\mu\nu M}=(B_{\mu\nu m}, {B_{\mu\nu}}^m)$. Therefore, only the subgroup $GL(5)\subset SO(5,5)$ is a manifest off-shell symmetry of the theory. Indices $M,N,\ldots =1,\ldots ,10$ denote fundamental or vector representation of $SO(5,5)$. Finally, there are $25$ scalar fields parametrizing the coset space $SO(5,5)/SO(5)\times SO(5)$. 

Fermionic fields, transforming under the local $SO(5)\times SO(5)$ symmetry, are symplectic Majorana-Weyl (SMW) spinors. Indices $\alpha,\ldots=1,\ldots,4$ and $\dot{\alpha},\ldots=\dot{1},\ldots,\dot{4}$ are respectively two sets of $SO(5)$ spinor indices in $SO(5)\times SO(5)$. Similarly, vector indices of the two $SO(5)$ factors are denoted by $a,\ldots=1,\ldots,5$ and $\dot{a},\ldots=\dot{1},\ldots,\dot{5}$. We use $\pm$ to indicate space-time chiralities of the spinors. Under the local $SO(5)\times SO(5)$ symmetry, the two sets of gravitini $\psi_{+\mu\alpha}$ and $\psi_{-\mu\dot{\alpha}}$ transform as $(\mathbf{4},\mathbf{1})$ and $(\mathbf{1},\mathbf{4})$ while the spin-$\frac{1}{2}$ fields $\chi_{+a\dot{\alpha}}$ and $\chi_{-\dot{a}\alpha}$ transform as $(\mathbf{5},\mathbf{4})$ and $(\mathbf{4},\mathbf{5})$, respectively.

In chiral spinor representation, the coset manifold $SO(5,5)/SO(5)\times SO(5)$ is described by a coset representative ${V_A}^{\alpha\dot{\beta}}$ transforming under the global $SO(5,5)$ and local $SO(5)\times SO(5)$ symmetries by left and right multiplications, respectively. The inverse elements ${(V^{-1})_{\alpha\dot{\beta}}}^A$ will be denoted by ${V^A}_{\alpha\dot{\beta}}$ satisfying the relations
\begin{equation}\label{VViProp}
{V_A}^{\alpha\dot{\beta}}{V^B}_{\alpha\dot{\beta}}=\delta^B_A\qquad \textrm{and}\qquad {V_A}^{\alpha\dot{\beta}}{V^A}_{\gamma\dot{\delta}}=\delta^{\alpha}_{\gamma}\delta^{\dot{\beta}}_{\dot{\delta}}\, .
\end{equation}
On the other hand, in vector representation, the coset representative is given by a $10\times10$ matrix ${\mathcal{V}_M}^{\underline{A}}=({\mathcal{V}_M}^{a}, {\mathcal{V}_M}^{\dot{a}})$ with $\underline{A}=(a,\dot{a})$. This is related to the coset representative in chiral spinor representation by the following relations
\begin{eqnarray}
{\mathcal{V}_M}^a&=&\frac{1}{16}V^{A\alpha\dot{\alpha}}(\Gamma_M)_{AB}{(\gamma^a)_{\alpha\dot{\alpha}}}^{\beta\dot{\beta}}{V^B}_{\beta\dot{\beta}},\label{VVrel1}\\{\mathcal{V}_M}^{\dot{a}}&=&-\frac{1}{16}V^{A\alpha\dot{\alpha}}(\Gamma_M)_{AB}{(\gamma^{\dot{a}})_{\alpha\dot{\alpha}}}^{\beta\dot{\beta}}{V^B}_{\beta\dot{\beta}}\, .\label{VVrel2}
\end{eqnarray}
In these equations, $(\Gamma_M)_{AB}$ and ${(\Gamma_{\ul{A}})_{\alpha\dot{\alpha}}}^{\beta\dot{\beta}}=({(\gamma_a)_{\alpha\dot{\alpha}}}^{\beta\dot{\beta}},{(\gamma_{\dot{a}})_{\alpha\dot{\alpha}}}^{\beta\dot{\beta}})$ are respectively $SO(5,5)$ gamma matrices in non-diagonal $\eta_{MN}$ and diagonal $\eta_{\ul{A}\ul{B}}$ bases.  
\\
\indent The inverse will be denoted by $\mathcal{V}^{M\underline{A}}$ satisfying the following relations 
\begin{equation}
\mathcal{V}^{Ma}{\mathcal{V}_M}^b=\delta^{ab},\qquad\mathcal{V}^{M\dot{a}}{\mathcal{V}_M}^{\dot{b}}=\delta^{\dot{a}\dot{b}},\qquad\mathcal{V}^{Ma}{\mathcal{V}_M}^{\dot{a}}=0
\end{equation}
and
\begin{equation}
{\mathcal{V}_M}^a\mathcal{V}^{Na}-{\mathcal{V}_M}^{\dot{a}}\mathcal{V}^{N\dot{a}}=\delta^N_M\, .\label{V_Vi_vec}
\end{equation}
In these equations, we have explicitly raised the $SO(5)\times SO(5)$ vector index $\underline{A}=(a,\dot{a})$ resulting in a minus sign in equation \eqref{V_Vi_vec}.

The most general gaugings of six-dimensional $N=(2,2)$ supergravity are described by the embedding tensor ${\Theta_A}^{MN}$ leading to the following covariant derivative
\begin{equation}\label{gauge_covariant_derivative}
D_\mu=\partial_\mu-gA^A_\mu\ {\Theta_A}^{MN}\boldsymbol{t}_{MN}
\end{equation}
where $g$ is a gauge coupling constant. The embedding tensor identifies generators $X_A={\Theta_A}^{MN}\boldsymbol{t}_{MN}$ of the gauge group $G_0\subset SO(5,5)$ with particular linear combinations of the $SO(5,5)$ generators $\boldsymbol{t}_{MN}$. Supersymmetry requires the embedding tensor to transform as $\mathbf{144}_c$ representation of $SO(5,5)$. Accordingly, ${\Theta_A}^{MN}$ can be parametrized in terms of a vector-spinor $\theta^{AM}$ of $SO(5,5)$ as 
\begin{equation}
{\Theta_A}^{MN}\ =\ -\theta^{B[M}(\Gamma^{N]})_{BA}\ \equiv \ \left(\Gamma^{[M}\theta^{N]}\right)_A
\end{equation}
with $\theta^{AM}$ subject to the constraint
\begin{equation}\label{MainLC}
(\Gamma_M)_{AB}\,\theta^{BM}\ =\ 0\, .
\end{equation}
\indent The $SO(5,5)$ generators in vector and spinor representations can be chosen as
\begin{equation}
{(\boldsymbol{t}_{MN})_P}^Q=4\eta_{P[M}\delta^Q_{N]}\qquad\text{and}\qquad {(\boldsymbol{t}_{MN})_A}^B\ =\ {(\Gamma_{MN})_A}^B
\end{equation}
with $\eta_{MN}$ being the off-diagonal $SO(5,5)$ invariant tensor given by
\begin{equation}\label{off-diag-eta}
\eta_{MN}\ =\ \eta^{MN}\ =\ \begin{pmatrix} 	0 & \mathds{1}_5 \\
							\mathds{1}_5 & 0    \end{pmatrix}
\end{equation}
and 
\begin{equation}
{(\Gamma_{MN})_A}^B=\frac{1}{2}\left[{(\Gamma_M)_A}^{C}{(\Gamma_N)_{C}}^B-{(\Gamma_N)_A}^{C}{(\Gamma_M)_{C}}^B\right].
\end{equation}
We also note that the notation $\mathds{1}_n$ denotes an $n\times n$ identity matrix. ${(\Gamma_M)_A}^{B}$ are chirally projected $SO(5,5)$ gamma matrices.
\\
\indent The corresponding gauge generators in these representations then take the forms
\begin{equation}\label{DefGaugeGen}
{(X_A)_M}^N=2\left(\Gamma_{M}\theta^{N}\right)_A+2\left(\Gamma^{N}\theta_{M}\right)_A\qquad\text{and}\qquad {(X_A)_B}^C= \left(\Gamma^{M}\theta^{N}\right)_A{(\Gamma_{MN})_B}^C\, .
\end{equation}
Furthermore, consistency requires the gauge generators to form a closed subalgebra of $SO(5,5)$ implying the quadratic constraint
\begin{equation}
\left[X_A,X_B\right]\ = \ -{(X_A)_B}^C\,X_C\, .
\end{equation} 
In terms of $\theta^{AM}$, the quadratic constraint reduces to the following two conditions
\begin{eqnarray}
\theta^{AM}\theta^{BN}\eta_{MN}&=&0,\label{QC1}\\ 
\theta^{AM}\theta^{B[N}(\Gamma^{P]})_{AB}&=&0\label{QC2}.
\end{eqnarray}  
It follows that any $\theta^{AM}\in\mathbf{144}_c$ satisfying this quadratic constraint defines a consistent gauging.
\\
\indent In this work, we are only interested in the classification of gauge groups under $GL(5)\subset SO(5,5)$ and domain wall solutions which only involve the metric and scalar fields. Therefore, we will, from now on, set all vector and tensor fields to zero with the bosonic Lagrangian of the maximal $N=(2,2)$ gauged supergravity given by
\begin{equation}
e^{-1}\mathcal{L}=\frac{1}{4}R-\frac{1}{16}P_{\mu}^{a\dot{a}}P^\mu_{a\dot{a}}-\mathbf{V}\, .\label{bosonic_L}
\end{equation}
We also need supersymmetry transformations of fermionic fields which, for vanishing fermions and vector/tensor fields, are given by
\begin{eqnarray}
\delta\psi_{+\mu\alpha}&=& D_\mu\epsilon_{+\alpha}+\frac{g}{4}\hat{\gamma}_\mu {T_\alpha}^{\dot{\beta}}\epsilon_{-\dot{\beta}},\label{1stSUSY}\\
\delta\psi_{-\mu\dot{\alpha}}&=& D_\mu\epsilon_{-\dot{\alpha}}-\frac{g}{4}\hat{\gamma}_\mu {T^{\beta}}_{\dot{\alpha}}\epsilon_{+\beta},\label{2ndSUSY}\\
\delta\chi_{+a\dot{\alpha}}&=&\frac{1}{4}P^\mu_{a\dot{a}}\hat{\gamma}_\mu{(\gamma^{\dot{a}})_{\dot{\alpha}}}^{\dot{\beta}}\epsilon_{-\dot{\beta}}+2g{(T_{a})^\beta}_{\dot{\alpha}}\epsilon_{+\beta}-\frac{g}{2}{T^{\alpha}}_{\dot{\alpha}}{(\gamma_a)_\alpha}^\beta\epsilon_{+\beta},\label{3rdSUSY}\\
\delta\chi_{-\dot{a}\alpha}&=&\frac{1}{4}P^\mu_{a\dot{a}}\hat{\gamma}_\mu{(\gamma^a)_\alpha}^\beta\epsilon_{+\beta}+2g{(T_{\dot{a}})_{\alpha}}^{\dot{\beta}}\epsilon_{-\dot{\beta}}+\frac{g}{2}{T_{\alpha}}^{\dot{\alpha}}{(\gamma_{\dot{a}})_{\dot{\alpha}}}^{\dot{\beta}}\epsilon_{-\dot{\beta}}.\label{4thtSUSY}
\end{eqnarray}
\indent The covariant derivatives of supersymmetry parameters, $\epsilon_{+\alpha}$ and $\epsilon_{-\dot{\alpha}}$, are defined by
\begin{eqnarray}
D_\mu\epsilon_{+\alpha}&=& \partial_\mu\epsilon_{+\alpha}+\frac{1}{4}{\omega_\mu}^{\nu\rho}\hat{\gamma}_{\nu\rho}\epsilon_{+\alpha}+\frac{1}{4}Q_\mu^{ab}{(\gamma_{ab})_\alpha}^\beta\epsilon_{+\beta},\label{CoDivEp+}\\
D_\mu\epsilon_{-\dot{\alpha}}&=& \partial_\mu\epsilon_{-\dot{\alpha}}+\frac{1}{4}{\omega_\mu}^{\nu\rho}\hat{\gamma}_{\nu\rho}\epsilon_{-\dot{\alpha}}+\frac{1}{4}Q_\mu^{\dot{a}\dot{b}}{(\gamma_{\dot{a}\dot{b}})_{\dot{\alpha}}}^{\dot{\beta}}\epsilon_{-\dot{\beta}}\label{CoDivEp-}
\end{eqnarray}
with $\hat{\gamma}_\mu=e_\mu^{\hat{\mu}}\hat{\gamma}_{\hat{\mu}}$. $\hat{\gamma}_{\hat{\mu}}$ are space-time gamma matrices, and for simplicity, we will suppress space-time spinor indices.

The scalar vielbein $P_{\mu}^{a\dot{a}}$ and $SO(5)\times SO(5)$ composite connections, $Q_\mu^{ab}$ and $Q_\mu^{\dot{a}\dot{b}}$, are given by
\begin{eqnarray}
P_{\mu}^{a\dot{a}}&=&\frac{1}{4}{(\gamma^a)}^{\alpha\beta}{(\gamma^{\dot{a}})}^{\dot{\alpha}\dot{\beta}}{V^A}_{\alpha\dot{\alpha}}\partial_\mu V_{A\beta\dot{\beta}},\label{PDef}\\
Q_{\mu}^{ab}&=&\frac{1}{8}{(\gamma^{ab})}^{\alpha\beta}\Omega^{\dot{\alpha}\dot{\beta}} {V^A}_{\alpha\dot{\alpha}}\partial_\mu V_{A\beta\dot{\beta}},\label{QuDef}\\
Q_{\mu}^{\dot{a}\dot{b}}&=&\frac{1}{8}\Omega^{\alpha\beta}{(\gamma^{\dot{a}\dot{b}})}^{\dot{\alpha}\dot{\beta}}{V^A}_{\alpha\dot{\alpha}}\partial_\mu V_{A\beta\dot{\beta}}\, .\label{QdDef}
\end{eqnarray}
In these equations, $\Omega^{\alpha\beta}$ and $\Omega^{\dot{\alpha}\dot{\beta}}$ are the $USp(4)$ symplectic forms that satisfy the following relations
\begin{equation}
\Omega_{\beta\alpha}\ =\ -\Omega_{\alpha\beta},\qquad \Omega^{\alpha\beta}\ =\ (\Omega_{\alpha\beta})^*,\qquad \Omega_{\alpha\beta}\Omega^{\beta\gamma}\ =\ -\delta^\gamma_\alpha
\end{equation}
and similarly for $\Omega_{\dot{\alpha}\dot{\beta}}$. We will use the explicit form of $\Omega_{\alpha\beta}$ and $\Omega_{\dot{\alpha}\dot{\beta}}$ given by
 \begin{equation}\label{USp(4)Omegas}
\Omega_{\alpha\beta}\ =\ \Omega_{\dot{\alpha}\dot{\beta}}\ =\ \mathds{1}_2\otimes i\sigma_2\, .
\end{equation}
We also note that the definitions \eqref{PDef}, \eqref{QuDef}, and \eqref{QdDef} can be derived from the following relation
\begin{equation}
{V^A}_{\alpha\dot{\alpha}}\partial_\mu V_{A\beta\dot{\beta}}=\frac{1}{4}P_{\mu}^{a\dot{a}}(\gamma_a)_{\alpha\beta}(\gamma_{\dot{a}})_{\dot{\alpha}\dot{\beta}}+\frac{1}{4}Q_{\mu}^{ab}(\gamma_{ab})_{\alpha\beta}\Omega_{\dot{\alpha}\dot{\beta}}+\frac{1}{4}Q_{\mu}^{\dot{a}\dot{b}}\Omega_{\alpha\beta}(\gamma_{\dot{a}\dot{b}})_{\dot{\alpha}\dot{\beta}}\, .
\end{equation}
\indent The scalar potential is given by
\begin{eqnarray}\label{scalarPot}
\mathbf{V}&=&\frac{g^2}{2}\theta^{AM}\theta^{BN}{\mathcal{V}_M}^a{\mathcal{V}_N}^b\left[{V_A}^{\alpha\dot{\alpha}}{(\gamma_a)_\alpha}^\beta{(\gamma_b)_\beta}^\gamma V_{B\gamma\dot{\alpha}}\right]\nonumber\\
&=&-\frac{g^2}{2}\left[T^{\alpha\dot{\alpha}}T_{\alpha\dot{\alpha}}-2(T^a)^{\alpha\dot{\alpha}}(T_a)_{\alpha\dot{\alpha}}\right]\label{scalarPot}
\end{eqnarray}
with the T-tensors defined by
\begin{equation}\label{TTenDef}
(T^a)^{\alpha\dot{\alpha}}={\mathcal{V}_M}^a\theta^{AM}{V_A}^{\alpha\dot{\alpha}},\qquad (T^{\dot{a}})^{\alpha\dot{\alpha}}=-{\mathcal{V}_M}^{\dot{a}}\theta^{AM}{V_A}^{\alpha\dot{\alpha}}
\end{equation}
and
\begin{equation}
T^{\alpha\dot{\alpha}}\equiv (T^a)^{\beta\dot{\alpha}}{(\gamma_a)_\beta}^\alpha=-(T^{\dot{a}})^{\alpha\dot{\beta}}{(\gamma_{\dot{a}})_{\dot{\beta}}}^{\dot{\alpha}}\, .
\end{equation}

\section{Gaugings of $N=(2,2)$ supergravity under $GL(5)$}\label{gauging}
In this section, we consider gaugings under $GL(5)\subset SO(5, 5)$. The embedding tensor $\theta^{AM}$ in $\mathbf{144}_c$ representation of $SO(5,5)$ decomposes under $GL(5)$ as
\begin{equation}\label{mainthetaDec}
\mathbf{144}_c\ \rightarrow\ \overline{\mathbf{5}}^{+3}\,\oplus\,\mathbf{5}^{+7}\,\oplus\,\mathbf{10}^{-1}\,\oplus\,\mathbf{15}^{-1}\,\oplus\,\mathbf{24}^{-5}\,\oplus\,\overline{\mathbf{40}}^{-1}\,\oplus\,\overline{\mathbf{45}}^{+3}\, .
\end{equation}
The $SO(5,5)$ spinor representation decomposes as 
\begin{equation}
\mathbf{16}_s\ \rightarrow\ \overline{\mathbf{5}}^{+3}\,\oplus\,\mathbf{10}^{-1}\,\oplus\,\mathbf{1}^{-5}\, . 
\end{equation}
Accordingly, the gauge generators can be written in terms of $X^m$, $X_{mn}$, and $X_\ast$ denoting respectively $\overline{\mathbf{5}}^{+3}$, $\mathbf{10}^{-1}$, and $\mathbf{1}^{-5}$ as
\begin{equation}\label{GaugeGenSplit}
X_A\ =\ \mathbb{T}_{Am}X^m+\mathbb{T}_{A}^{mn}X_{mn}+\mathbb{T}_{A\ast}X_\ast\, .
\end{equation}
The decomposition matrices $\mathbb{T}_{Am}$, $\mathbb{T}_{A}^{mn}$, and $\mathbb{T}_{A\ast}$ are given in the appendix.
\\
\indent By the decomposition of $SO(5,5)$ vector representation
\begin{equation}\label{VecDec}
\mathbf{10}\ \rightarrow\ \mathbf{5}^{+2}\,\oplus\,\overline{\mathbf{5}}^{-2},
\end{equation} 
we can write the embedding tensor as
\begin{equation}
\theta^{AM}=(\theta^{Am},\theta^{A}_m).
\end{equation}
The two components $\theta^{Am}$ and $\theta^{A}_m$ contain the following irreducible $GL(5)$ representations
\begin{eqnarray}
\theta^{Am}&:&\qquad \overline{\mathbf{5}}^{+3}\,\oplus\,\mathbf{10}^{-1}\,\oplus\,\mathbf{24}^{-5}\,\oplus\,\overline{\mathbf{40}}^{-1},\label{splitthetaDec1}\\ 
\theta^{A}_m&:&\qquad \overline{\mathbf{5}}^{+3}\,\oplus\,\mathbf{5}^{+7}\,\oplus\,\mathbf{10}^{-1}\,\oplus\,\mathbf{15}^{-1}\,\oplus\,\overline{\mathbf{45}}^{+3}\, .\label{splitthetaDec2}
\end{eqnarray}
As pointed out in \cite{6D_Max_Gauging}, gaugings triggered by $\theta^{Am}$ are called electric gaugings in the sense that only electric two-forms participate in the gauged theory while gaugings triggered by $\theta^A_m$ are called magnetic gaugings involving magnetic two-forms together with additional three-form tensor fields. The decompositions in \eqref{splitthetaDec1} and \eqref{splitthetaDec2} imply that gaugings in $\mathbf{24}^{-5}\,\oplus\,\overline{\mathbf{40}}^{-1}$ and $\mathbf{5}^{+7}\,\oplus\,\mathbf{15}^{-1}\,\oplus\,\overline{\mathbf{45}}^{+3}$ representations are respectively purely electric and purely magnetic while those in $\overline{\mathbf{5}}^{+3}\,\oplus\,\mathbf{10}^{-1}$ representations correspond to dyonic gaugings. Many possible dyonic gaugings can also arise from combinations of various electric and magnetic components. Finally, we note that the quadratic constraint \eqref{QC1} is automatically satisfied for purely electric or purely magnetic gaugings that involve only $\theta^{Am}$ or $\theta^{A}_m$ components.
\\
\indent In accord with \eqref{splitthetaDec1} and \eqref{splitthetaDec2}, we can parametrize the embedding tensor as, see \cite{6D_DW_I} for more detail,
\begin{eqnarray}
\theta^{Am}&=&\mathbb{T}^{An}{S_n}^m+\mathbb{T}^{A}_{np}\left(U^{np,m} + \frac{1}{3\sqrt{2}}\varepsilon^{mnpqr}Z_{qr}\right)-\frac{2\sqrt{2}}{5}\mathbb{T}^{A}_{\ast}J^m,\label{ExpliTheta1}\\
 \theta^{A}_m&=&\mathbb{T}^{An}(Y_{nm} + Z_{nm})+\mathbb{T}^{A}_{np}(W^{np}_{m}+J^{[n}\delta^{p]}_m)+\mathbb{T}^{A}_\ast K_m\, .\label{ExpliTheta2}
\end{eqnarray}
Matrices $\mathbb{T}^{Am}$, $\mathbb{T}^{A}_{mn}$, and $\mathbb{T}^{A}_{\ast}$ are inverses of the decomposition matrices $\mathbb{T}_{Am}$, $\mathbb{T}_{A}^{mn}$, and $\mathbb{T}_{A\ast}$ given by complex conjugations, $\mathbb{T}^A=(\mathbb{T}_A)^{-1}=(\mathbb{T}_A)^*$. 
\\
\indent We now look at various components in more detail. The first representation $\mathbf{24}^{-5}$ is described by a traceless $5\times5$ matrix ${S_n}^m$ with ${S_m}^m=0$. The tensor $U^{np,m}=U^{[np],m}$ satisfying $U^{[np,m]}=0$ corresponds to $\overline{\mathbf{40}}^{-1}$ representation. The symmetric tensor $Y_{mn}=Y_{(mn)}$ and antisymmetric one $Z_{mn}=Z_{[mn]}$ respectively denote $\mathbf{15}^{-1}$ and $\mathbf{10}^{-1}$ representations. The $\overline{\mathbf{45}}^{+3}$ representation is written as the tensor $W^{np}_m=W^{[np]}_m$ satisfying $W^{nm}_m=0$. Finally, the last two $\overline{\mathbf{5}}^{+3}$ and $\mathbf{5}^{+7}$ representations are parametrized by two $GL(5)$ vectors $J^m$ and $K_m$, respectively. The general structure of gaugings under $GL(5)\subset SO(5,5)$ is shown in Table \ref{Tab1} taken from \cite{6D_Max_Gauging}. The left column represents the sixteen vector fields in $GL(5)$ representations while the top row corresponds to the decomposition of $SO(5,5)$ generators under $GL(5)$. The table shows the couplings between $SO(5,5)$ generators and vector fields by various components of the embedding tensor. 

The gaugings from the components $\mathbf{15}^{-1}$ and $\overline{\mathbf{40}}^{-1}$ have been extensively studied in \cite{6D_DW_I}. In the present work, we will focus on the remaining representations and examples of possible combinations shown in Table \ref{Tab1}. In the following, we will determine an explicit form of the aforementioned $GL(5)$ tensors by imposing the quadratic constraints \eqref{QC1} and \eqref{QC2} on the embedding tensor under consideration.  

\begin{table}[h!]
\begin{center}
\begin{tabular}{| c | c c c c|}
\hline
${\Theta_A}^{MN}$ & $\overline{\mathbf{10}}^{-4}$ & $\mathbf{1}^{0}$ & $\mathbf{24}^{0}$ & $\mathbf{10}^{+4}$\\ \hline
$\mathbf{5}^{-3}$ & $\mathbf{5}^{+7}$ & $\overline{\mathbf{5}}^{+3}$ & $(\overline{\mathbf{5}}+\overline{\mathbf{45}})^{+3}$ & $(\mathbf{10}+\overline{\mathbf{40}})^{-1}$\\
$\overline{\mathbf{10}}^{+1}$ & $(\overline{\mathbf{5}}+\overline{\mathbf{45}})^{+3}$ & $\mathbf{10}^{-1}$ & $(\mathbf{10}+\mathbf{15}+\overline{\mathbf{40}})^{-1}$ & $\mathbf{24}^{-5}$\\
$\mathbf{1}^{+5}$ & $\mathbf{10}^{-1}$ & & $\mathbf{24}^{-5}$ &\\ \hline
\end{tabular}
\caption{Gauge couplings between the sixteen vector fields and $SO(5,5)$ generators from various $GL(5)$ components of the embedding tensor.}\label{Tab1}
\end{center}
\end{table}

\subsection{Gaugings in $\mathbf{24}^{-5}$ representation}\label{24repSec}
We begin with gauge groups arising from the embedding tensor in $\mathbf{24}^{-5}$ representation. Gaugings in this representation are purely electric and triggered by
\begin{equation}
\theta^{Am}\ = \ \mathbb{T}^{An}{S_n}^m\, .\label{theta_24_5}
\end{equation}
With $\theta^{A}_m=0$, the embedding tensor $\theta^{AM}=(\,\mathbb{T}^{An}{S_n}^m\,,\,0\,)$ automatically satisfies the quadratic constraint. Therefore, every traceless $5\times5$ matrix ${S_m}^n$ defines a viable gauge group generated by the following gauge generators
\begin{equation}\label{24gaugeGen}
X^m\ =\ 0,\qquad X_{mn}\ =\ {S_m}^p\boldsymbol{s}_{pn}-{S_n}^p\boldsymbol{s}_{pm},\qquad
X_\ast\ =-\sqrt{2}{S_m}^n{\boldsymbol{\tau}^m}_n\, .
\end{equation}
${\boldsymbol{\tau}^m}_n$ are $SL(5)$ generators defined as
\begin{equation}\label{DefSL5Op}
{\boldsymbol{\tau}^m}_n\ =\ {\boldsymbol{t}^m}_n-\frac{1}{5}\,\boldsymbol{d}\,\delta^m_n
\end{equation}
with ${\boldsymbol{\tau}^m}_m=0$ and
\begin{equation} \boldsymbol{d}={\boldsymbol{t}^m}_m={\boldsymbol{t}^1}_1+{\boldsymbol{t}^2}_2+{\boldsymbol{t}^3}_3
+{\boldsymbol{t}^4}_4+{\boldsymbol{t}^5}_5\label{SO(1,1)Gen}
\end{equation} 
being the $SO(1,1)\sim \mathbb{R}^+$ generator in $GL(5)\sim \mathbb{R}^+\times SL(5)$. We also use $\boldsymbol{s}_{mn}=\boldsymbol{t}_{mn}$ to denote the generators corresponding to shift symmetries on the scalar fields, see more detail in the appendix.

Commutation relations between the gauge generators read
\begin{equation}\label{24algebra}
[X_{mn},X_{pq}]\ =\ 0\qquad\text{ and }\qquad [X_{mn},X_\ast]\ =\ {(X_\ast)_{mn}}^{pq}X_{pq}
\end{equation}
where the generators with two antisymmetric pairs of $GL(5)$ vector indices are defined as 
\begin{equation}
{(X_A)_{mn}}^{pq}=2{(X_A)_{[m}}^{[p}\delta^{q]}_{n]}\, .
\end{equation}
From equation \eqref{24algebra}, we readily see that the generators $X_{mn}$ form a translational group and transform non-trivially under a one-dimensional group generated by $X_\ast$, a particular linear combination of $SL(5)$ generators.

In vector or fundamental representation of $SL(5)$, $X_\ast$ generator takes the form 
\begin{equation}
{(X_\ast)_m}^n=-2\sqrt{2}{S_m}^n\, .
\end{equation}
If ${S_m}^n$ is antisymmetric, $X_\ast$ will also be antisymmetric and generates a compact $SO(2)$ group. On the other hand, for symmetric ${S_m}^n$, $X_\ast$ generates a non-compact $SO(1,1)$ group. The resulting gauge groups then take the form of
\begin{equation}
SO(2)\ltimes \mathbb{R}^n \qquad \textrm{and}\qquad SO(1,1)\ltimes \mathbb{R}^n
\end{equation}
for $3\leq n\leq 10$. The values of $n$ depend on the choices of ${S_m}^n$. It has been pointed out in \cite{6D_Max_Gauging} that this type of gaugings is related to Scherk-Schwarz reductions from seven-dimensional gauged supergravity, and gaugings in $\mathbf{24}^{-1}$ representation correspond to choosing a generator from the seven-dimensional symmetry group $SL(5)$, see also \cite{maximal_SUGRA_deWit} for a general discussion on Scherk-Schwarz reductions and gauged supergravities. 

We end this case by giving an explicit example with $X_\ast$ antisymmetric. By choosing
\begin{equation}\label{Schoices}
{S_m}^n=\begin{small}\begin{pmatrix} 0 & \kappa & 0 & 0 & 0\\ -\kappa & 0 & 0 & 0 & 0\\ 0 & 0 & 0 & 0 & 0\\ 0 & 0 & 0 & 0 & -\lambda\\ 0 & 0 & 0 & \lambda & 0 \end{pmatrix}\end{small},
\end{equation}
we find the following non-vanishing gauge generators
\begin{eqnarray}
X_{ix}&=&{S_{i}}^{j}\boldsymbol{s}_{jx}-{S_{x}}^{y}\boldsymbol{s}_{yi},\qquad X_{i3}\ =\ {S_{i}}^{j}\boldsymbol{s}_{j3},\qquad X_{x3}\ =\ {S_{x}}^{y}\boldsymbol{s}_{y3},\nonumber\\
 X_\ast&=&-\sqrt{2}(\kappa{\boldsymbol{\tau}^1}_2-\kappa{\boldsymbol{\tau}^2}_1-\lambda{\boldsymbol{\tau}^4}_5+\lambda{\boldsymbol{\tau}^5}_4)\label{SgaugeGen}
\end{eqnarray}
in which $i,j=1,2$ and $x,y=4,5$. It can be straightforwardly verified that these gauge generators satisfy the commutation relations given in \eqref{24algebra}. The resulting gauge group is of the form
\begin{equation}
G_0=SO(2)\ltimes \mathbb{R}^{8}_{\boldsymbol{s}}
\end{equation}
in which $SO(2)$ is generated by $X_\ast$, and the shift translational group $\mathbb{R}^{8}_{\boldsymbol{s}}$ is generated by $X_{ix}$, $X_{i3}$, and $X_{x3}$.

\subsection{Gaugings in $\overline{\mathbf{45}}^{+3}$ representation}\label{45repSec}
We now consider the embedding tensor in $\overline{\mathbf{45}}^{+3}$ representation. Gaugings in this case are purely magnetic and related to reductions of eleven-dimensional supergravity on twisted tori as pointed out in \cite{6D_Max_Gauging}. The linear constraint \eqref{MainLC} and the quadratic constraint \eqref{QC1} are both satisfied if we parametrize the embedding tensor as 
\begin{equation}\label{45theta}
\theta^{AM}=(\,0\,,\,\mathbb{T}^{A}_{np} W^{np}_m\,).
\end{equation}
In terms of $W^{np}_m$, the constraint \eqref{QC2} reduces to
\begin{equation}\label{45QC}
W^{mn}_r\,W^{pq}_{[s}\,\varepsilon_{t]mnpq}\ =\ 0\, .
\end{equation}
As noted in \cite{6D_Max_Gauging}, this constraint is the duality of a similar condition in gaugings from $\overline{\mathbf{40}}^{-1}$ representation ($U^{mn,r}U^{pq,s}\varepsilon_{mnpqt}\ = \ 0$), considered in \cite{6D_DW_I}. To solve this condition, we then follow the same procedure as in $\overline{\mathbf{40}}^{-1}$ representation. We first write $W^{np}_m$ in the form of
\begin{equation}\label{45Wansatz}
W^{np}_m\ =\ 2v^{[n}{u_m}^{p]}
\end{equation}
and impose the condition ${u_m}^{m}=0$ in order to satisfy the traceless condition $W^{nm}_m=0$. This form of $W^{np}_m$ is sufficient to solve the condition \eqref{45QC}. 

To give an explicit example, we will choose a basis in which $v^m=\delta^m_5$ and ${u_5}^{m}={u_m}^{5}=0$. As a result, consistent gaugings satisfying the linear and quadratic constraints in $\overline{\mathbf{45}}^{+3}$ representation are now parametrized by a traceless $4\times 4$ matrix ${u_i}^j$ with $i,j=1,...,4$ and ${u_i}^{i}=0$. In this case, the gauge generators turn out to be $X_\ast=X_{i5}=0$ and the remaining non-vanishing generators given by
\begin{equation}
X_{ij}\ = \ \sqrt{2}\varepsilon_{ijkl}{u_m}^k\boldsymbol{h}^{lm},\qquad X^i\ = -2{u_k}^i{\boldsymbol{\tau}^k}_5,\qquad X^5\ = 2{u_k}^j{\boldsymbol{\tau}^k}_j\, .
\end{equation}
We recall that $\boldsymbol{h}^{mn}=\boldsymbol{t}^{mn}$ are $SO(5,5)$ generators associated with the hidden symmetries that do not constitute symmetries of the action. Commutation relations between the gauge generators read
\begin{eqnarray}
\left[X_{ij},X_{kl}\right]&=&\left[X_{ij},X^{k}\right]\ = \ \left[X^i,X^j\right]\ = \ 0,\nonumber\\
\left[X^5,X_{ij}\right]&=&{(X^5)_{ij}}^{kl}X_{kl},\qquad\quad \left[X^5,X^i\right]={(X^5)_{j}}^{i}X^j.
\end{eqnarray}
As seen from these relations, $X_{ij}$ generate a six-dimensional translational group $\mathbb{R}^6_{\boldsymbol{h}}$ associated with the hidden symmetries while $X^i$  generate another translational group $\mathbb{R}^4$ commuting with $\mathbb{R}^6_{\boldsymbol{h}}$. The $X^5$ generator takes the same form as $X_\ast$ in $\mathbf{24}^{-1}$ representation. We can similarly use the unbroken $SL(4)$ symmetry to fix ${u_i}^j$ in the form of
\begin{equation}\label{uchoices}
\small
{u_i}^j=\begin{pmatrix} 0 & \kappa & 0 & 0\\ \pm\kappa & 0 & 0 & 0\\  0 & 0 & 0 & -\lambda\\ 0 & 0 & \mp\lambda & 0 \end{pmatrix}\, .
\end{equation}
With this explicit form of ${u_i}^j$, non-vanishing gauge generators are now given by
\begin{eqnarray}
X_{\bar{i}\bar{x}}&=&-\sqrt{2}\varepsilon_{\bar{i}\bar{j}}\varepsilon_{\bar{x}\bar{y}}({u_{\bar{k}}}^{\bar{j}}\boldsymbol{h}^{\bar{y}\bar{k}}-{u_{\bar{z}}}^{\bar{y}}\boldsymbol{h}^{\bar{j}\bar{z}}),\qquad X^i\ = -2{u_k}^i{\boldsymbol{\tau}^k}_5,\nonumber\\ X^5&=&2(\kappa{\boldsymbol{\tau}^1}_2\pm\kappa{\boldsymbol{\tau}^2}_1-\lambda{\boldsymbol{\tau}^3}_4\mp\lambda{\boldsymbol{\tau}^4}_3)
\end{eqnarray}
where $i=(\bar{i},\bar{x})$ with $\bar{i},\bar{j},...=1,2$ and $\bar{x},\bar{y},...=3,4$. Therefore, for arbitrary values of $\kappa$ and $\lambda$, the corresponding gauge groups are given by
\begin{equation}\label{45RepSO11GG}
G_0=SO(1,1)\ltimes\left(\mathbb{R}^4\times\mathbb{R}^4_{\boldsymbol{h}}\right)\sim CSO(1,1,2)\ltimes\mathbb{R}^4_{\boldsymbol{h}}
\end{equation}
or
\begin{equation}\label{45RepSO2GG}
G_0=SO(2)\ltimes\left(\mathbb{R}^4\times\mathbb{R}^4_{\boldsymbol{h}}\right)\sim CSO(2,0,2)\ltimes\mathbb{R}^4_{\boldsymbol{h}}
\end{equation}
depending on the choices of the signs for $\kappa$ and $\lambda$, making ${u_i}^j$ in \eqref{uchoices} respectively symmetric or antisymmetric. 

\subsection{Gaugings in $\mathbf{5}^{+7}$ representation}\label{5repSec}
In this section, we look at the smallest representation giving rise to purely magnetic gaugings. With $\theta^{Am}=0$, we parametrize the embedding tensor, in this case, by a $GL(5)$ vector $K_m$ as
\begin{equation}
\theta^A_m\ =\ \mathbb{T}^{A}_\ast K_m\, .
\end{equation}
Gaugings triggered by $\theta^{AM}=(\,0\,,\,\mathbb{T}^{A}_\ast K_m\,)$ automatically satisfy the linear and quadratic constraints. As pointed out in \cite{6D_Max_Gauging}, these gaugings might correspond to reductions from eleven dimensions with non-trivial four-form fluxes.

Only gauge generators $X^m$ are non-vanishing in this representation. They are explicitly given in terms of the generators of the hidden symmetry by
\begin{equation}
X^m\ =\ 5\sqrt{2}K_n\boldsymbol{h}^{nm}\, .
\end{equation}
These generators commute with each other
\begin{equation}
\left[X^m,X^n\right]\ =\ 0\, .
\end{equation}
However, due to the antisymmetric property of $\boldsymbol{h}^{nm}$, the gauge generators satisfy the condition $K_mX^m=0$. Therefore, only four generators are linearly independent. The resulting gauge group in $\mathbf{5}^{+7}$ representation is then given by a four-dimensional translational group $\mathbb{R}^4_{\boldsymbol{h}}$ associated with the hidden symmetries.

To explicitly parametrize this $\mathbb{R}^4_{\boldsymbol{h}}$ gauge group, we will use the $SL(5)$ symmetry to fix $K_m=\kappa\,\delta^5_m$ with $\kappa\in\mathbb{R}$. In this case, the gauge group is generated by 
\begin{equation}
X^i\ =\ 5\sqrt{2}\kappa\,\boldsymbol{h}^{5i}
\end{equation}
for $i=1,2,3,4$. 

\subsection{Gaugings in $\overline{\mathbf{5}}^{+3}$ representation}\label{5bsection}
We now consider dyonic gaugings involving both electric and magnetic components of the embedding tensor. We begin with gaugings in $\overline{\mathbf{5}}^{+3}$ representation. The embedding tensor in this representation, satisfying the linear constraint \eqref{MainLC}, is given by $\theta^{AM}=(\theta^{Am},\theta^{A}_m)$ with
\begin{equation}\label{5btheta}
\theta^{Am}\ = -\frac{2\sqrt{2}}{5}\mathbb{T}^{A}_\ast J^m\qquad\text{ and }\qquad \theta^{A}_m\ = \ \mathbb{T}^{A}_{np} J^{[n}\delta^{p]}_m\, .
\end{equation}
This form of the embedding tensor automatically satisfies the quadratic constraints given in \eqref{QC1} and \eqref{QC2}. Therefore, any vectors $J^m$ define consistent gaugings in $\overline{\mathbf{5}}^{+3}$ representation.

In this case, we find non-vanishing gauge generators given by
\begin{equation}
X^m=3{J^n\boldsymbol{\tau}^m}_n+\frac{8}{5}J^m\boldsymbol{d}\qquad\text{ and }\qquad
X_{mn}=-\frac{1}{\sqrt{2}}\varepsilon_{mnpqr}J^p\boldsymbol{h}^{qr}
\end{equation}
with the following commutation relations 
\begin{equation}
\left[X_{mn},X_{pq}\right]=0,\qquad \left[X^m,X_{np}\right]={(X^m)_{np}}^{qr}X_{qr},\qquad \left[X^m,X^n\right]=3{(X^{[m})_p}^{n]}X^p\, .\label{dyon5algebras}
\end{equation}
To write down the explicit form of gauge generators, we can use $SL(5)$ symmetry to fix $J^m=\kappa\,\delta^m_5$ as in the previous case. With this choice, non-vanishing gauge generators are given by
\begin{eqnarray}
X_{ij}&=&-\frac{\kappa}{\sqrt{2}}\varepsilon_{ijkl}\boldsymbol{h}^{kl},\qquad X^i\ =\ 3\kappa\,{\boldsymbol{\tau}^i}_5,\nonumber\\
 X^5&=&\kappa\left({\boldsymbol{\tau}^1}_1+{\boldsymbol{\tau}^2}_2+{\boldsymbol{\tau}^3}_3+{\boldsymbol{\tau}^4}_4+4\,{\boldsymbol{\tau}^5}_5\right)
\end{eqnarray}
with $i,j,...=1,...,4$. In vector representation of $SL(5)$, the generator $X^5$ takes the form
\begin{equation}\label{nonComJ}
{(X^5)_m}^n=2\kappa\, \text{diag}(1,1,1,1,4).
\end{equation}
Commutation relations between these gauge generators become
\begin{eqnarray}
\left[X_{ij},X_{kl}\right]&=&\left[X^i,X^j\right]\ =\ \left[X^i,X_{jk}\right]\ =\ 0,\nonumber\\
\left[X^5,X^i\right]&=&-3{(X^5)_j}^iX^j,\quad\qquad \left[X^5,X_{ij}\right]\ =\ {(X^5)_{ij}}^{kl}X_{kl}\, .\label{Fixeddyon5algebra}
\end{eqnarray}
These imply the gauge group of the form
\begin{equation}\label{nonComJGG}
G_0=SO(1,1)\ltimes\left(\mathbb{R}^4\times\mathbb{R}^6_{\boldsymbol{h}}\right)\sim CSO(1,1,2)\ltimes\mathbb{R}^6_{\boldsymbol{h}}
\end{equation}
in which the non-compact factor $SO(1,1)$ is generated by $X^5$. 

\subsection{Gaugings in $\mathbf{10}^{-1}$ representation}\label{10Sec}
In addition to gaugings from $\overline{\mathbf{5}}^{+3}$ representation, gaugings in $\mathbf{10}^{-1}$ representation also require both $\theta^{Am}$ and $\theta^A_m$ to be non-vanishing in order to satisfy the linear constraint. Thus, gaugings in this representation are dyonic and triggered by
\begin{equation}
\theta^{Am}\ =\ \frac{1}{3\sqrt{2}}\,\mathbb{T}^{A}_{np}\,\varepsilon^{mnpqr}Z_{qr}\qquad\text{ and }\qquad \theta^A_m\ = \ \mathbb{T}^{An}Z_{nm}
\end{equation}
with $Z_{mn}=Z_{[mn]}$. With this embedding tensor, the quadratic constraints in \eqref{QC1} and \eqref{QC2} reduce to
\begin{equation}\label{10QC}
Z_{mn}\, Z_{pq}\, \varepsilon^{mnpqr}\ = \ 0\, .
\end{equation}
As pointed in \cite{6D_Max_Gauging}, this condition can be solved by $Z_{mn}$ of the form
\begin{equation}\label{ZExform}
Z_{mn}=u_{[m}v_{n]}
\end{equation}
with $u_m$ and $v_n$ being arbitrary $GL(5)$ vectors.

In this case, the corresponding gauge generators are given by
\begin{eqnarray}
X_\ast&=&-\sqrt{2}u_{[m}v_{n]}\boldsymbol{h}^{mn},\label{10repgaugeGen1}\\ 
X^m&=&-\frac{\sqrt{2}}{3}\varepsilon^{mnpqr}u_{[n}v_{p]}\boldsymbol{s}_{qr},\label{10repgaugeGen2}\\ 
X_{mn}&=&\frac{1}{3}\left(u_{[m}v_{p]}{\boldsymbol{\tau}^p}_n-u_{[n}v_{p]}{\boldsymbol{\tau}^p}_m\right)+\frac{4}{5}u_{[m}v_{n]}\boldsymbol{d}\label{10repgaugeGen3}
\end{eqnarray}
with the commutation relations 
\begin{eqnarray}
\left[X_\ast,X^m\right]&=&\left[X^m,X^n\right]\ = \ 0,\nonumber\\
\left[X_{mn},X_\ast\right]&=&{(X_{mp})_n}^pX_\ast,\qquad \left[X^m,X_{np}\right]\ = \ {2(X_{np})_q}^mX^q,\nonumber\\
\left[X_{mn},X_{pq}\right]&=&\frac{1}{4}\left[{(X_{mn})_{pq}}^{rs}-{(X_{pq})_{mn}}^{rs}\right]X_{rs}\, .
\end{eqnarray}
\indent To determine the explicit form of possible gauge groups, we repeat the same procedure as in the previous cases by considering a particular parametrization of the two vectors $u_m$ and $v_m$ in the form of 
\begin{equation}
u_m=(0,0,0,\kappa_1,\kappa_2)\quad\text{ and }\quad v_m=(0,0,0,\lambda_1,\lambda_2).
\end{equation}
The gauge generators in this case become
\begin{eqnarray}
X_\ast&=&\sqrt{2}(\kappa_2\lambda_1-\kappa_1\lambda_2)\boldsymbol{h}^{45},\\
X^i&=&\frac{\sqrt{2}}{3}(\kappa_2\lambda_1-\kappa_1\lambda_2)\varepsilon^{ijk}\boldsymbol{s}_{jk},\\ 
X_{ix}&=&\frac{1}{6}(u_yv_x-u_xv_y){\boldsymbol{\tau}^y}_i,\\
X_{45}&=&\frac{1}{6}(\kappa_1\lambda_2-\kappa_2\lambda_1)\left({\boldsymbol{\tau}^4}_4+{\boldsymbol{\tau}^5}_5+\frac{12}{5}\boldsymbol{d}\right)
\end{eqnarray}
with $m=(i,x)$ for $i,j,...=1,2,3$ and $x,y,...=4,5$. The commutation relations read 
\begin{eqnarray}
\left[X_\ast,X^i\right]&=&\left[X_\ast,X_{ix}\right]\ = \ \left[X^i,X^j\right]\ = \ \left[X^i,X_{jx}\right]\ = \ \left[X_{ix},X_{jy}\right]\ =\ 0,\qquad\nonumber\\
\left[X_{45},X_\ast\right]&=&{(X_{45})_i}^iX_\ast,\qquad\left[X_{45},X^{i}\right]\ =\ -2{(X_{45})_j}^iX^j,\nonumber\\
\left[X_{45},X_{ix}\right]&=&\frac{2}{5}{(X_{45})_{ix}}^{jy}X_{jy}
\end{eqnarray}
giving rise to the gauge group of the form
\begin{equation}\label{case1rep10GG}
G_0\ =\ SO(1,1)\ltimes(\mathbb{R}^6\times\mathbb{R}^{3}_{\boldsymbol{s}}\times\mathbb{R}_{\boldsymbol{h}})\sim CSO(1,1,3)\ltimes(\mathbb{R}^{3}_{\boldsymbol{s}}\times\mathbb{R}_{\boldsymbol{h}}).
\end{equation}
The non-compact subgroup $SO(1,1)$ is generated by the gauge generator $X_{45}$ whose explicit form in $GL(5)$ vector representation is given by
\begin{equation}\label{case1SO11Gen}
{(X_{45})_m}^n=\frac{1}{3}(\kappa_1\lambda_2-\kappa_2\lambda_1)\,\text{diag}(2,2,2,3,3).
\end{equation}
The three commuting translational groups $\mathbb{R}^6$, $\mathbb{R}^{3}_{\boldsymbol{s}}$, and $\mathbb{R}_{\boldsymbol{h}}$ are respectively generated by $X_{ix}$, $X^{i}$, and $X_\ast$. 

\subsection{Gaugings in $(\overline{\mathbf{5}}+\overline{\mathbf{45}})^{+3}$ representation}\label{5+45repSec}
We now consider dyonic gaugings arising from combining two components of the embedding tensor in $\overline{\mathbf{5}}^{+3}$ and $\overline{\mathbf{45}}^{+3}$ representations. These gaugings are also dyonic since the embedding tensor in $\overline{\mathbf{5}}^{+3}$ representation contains both electric and magnetic parts. The linear constraint \eqref{MainLC} requires these components to take the form
\begin{equation}\label{5+45theta}
\theta^{Am}\ = -\frac{2\sqrt{2}}{5}\mathbb{T}^{A}_\ast J^m\qquad\text{ and }\qquad \theta^{A}_m\ = \ \mathbb{T}^{A}_{np}( J^{[n}\delta^{p]}_m+W^{np}_m).
\end{equation}
This is just a trivial combination between the embedding tensor from each representation given in \eqref{45theta} and \eqref{5btheta}.
\\ 
\indent With this form of the embedding tensor, the quadratic constraints \eqref{QC1} and \eqref{QC2} become
\begin{eqnarray}
W^{mn}_pJ^p&=&0,\label{5+45QC1}\\ W^{qr}_mW^{st}_{[n}\varepsilon_{p]qrst}+\frac{3}{2}W^{qr}_mJ^s\varepsilon_{npqrs}&=&0\, .\label{5+45QC2}
\end{eqnarray}
To solve these conditions, we first fix the explicit form of the $GL(5)$ vector $J^m=\kappa\delta^m_5$ with $\kappa\in\mathbb{R}$, leading to a split of a $GL(5)$ index $m=(i,5)$ with $i,j,...=1,2,3,4$ as in section \ref{5bsection}. With $J^m$ of this form, the condition \eqref{5+45QC1} implies $W^{mn}_5=0$, so only $W^{ij}_k$ and $W^{i5}_j$ are non-vanishing. Note also that the condition $W^{mn}_n=0$ imposes a traceless condition $W^{ij}_j=0$.

With only $W^{ij}_k$, $W^{i5}_j$, and $J^5=\kappa$ non-vanishing, the condition \eqref{5+45QC2} splits into
\begin{eqnarray}
W^{kl}_iW^{mn}_{j}\varepsilon_{klmn}&=&0,\label{5+45QC2.1}\\ 2(W^{lm}_iW^{n5}_{[j}+W^{l5}_iW^{mn}_{[j})\varepsilon_{k]lmn}+\frac{3}{2}\kappa W^{lm}_i\varepsilon_{jklm}&=&0\, .\label{5+45QC2.2}
\end{eqnarray}
The first condition takes the same form as the quadratic constraint \eqref{45QC} in $\overline{\mathbf{45}}^{+3}$ representation and can be similarly solved by choosing 
\begin{equation}\label{5+45Wansatz1}
W^{ij}_k\ =\ 2v^{[i}{w_k}^{j]}
\end{equation}
with a $GL(4)$ vector $v^i$ and a traceless $4\times 4$ matrix ${w_k}^j$. 

At this point, we can further use $SL(4)\subset GL(5)$ symmetry to rotate $v^i$ such that $v^i=\lambda\delta^i_4$. We will also split the index $i=(x,4)$ with $x=1,2,3$. Moreover, the traceless condition $W^{ij}_j=0$ requires that ${w_4}^i=0$. For simplicity, we will also set ${w_i}^4=0$ since these components do not appear in the resulting embedding tensor. The remaining components ${w_x}^y$ can be described by a traceless $3\times 3$ matrix satisfying ${w_x}^x=0$. Among various components of $W^{ij}_k=(W^{xy}_z,\ W^{ij}_4,\ W^{4y}_x)$, only the last components are non-vanishing and given by
\begin{equation}
W^{4y}_x=\lambda {w_x}^y\, .
\end{equation}
With all these, the condition \eqref{5+45QC2.2} splits into
\begin{eqnarray}
\lambda W^{x5}_4{w_z}^t\varepsilon_{xyt}&=&0,\label{5+45QC2.2.1}\\
2\left[{w_x}^tW^{5a}_{[y}+W^{5a}_x{w_{[y}}^t\right]\varepsilon_{z]ta}+\frac{3}{2}\kappa{w_x}^t\varepsilon_{yzt}&=&0\label{5+45QC2.2.2}.
\end{eqnarray}
The first condition can be solved by setting $W^{x5}_4=0$. The remaining component $W^{x5}_y$ can be written in terms of a $3\times 3$ matrix ${u_y}^x$ as $W^{5x}_y=\tau {u_y}^x$ with $\tau\in\mathbb{R}$. The condition \eqref{5+45QC2.2.2} then gives rise to
\begin{equation}
2\tau\left[({w_x}^t{u_y}^a+{w_y}^t{u_x}^a)\varepsilon_{zta}-({w_x}^t{u_z}^a+{w_z}^t{u_x}^a)\varepsilon_{yta}\right]
+3\kappa{w_x}^t\varepsilon_{yzt}=0\, .
\end{equation}
With the Schouten identity, ${w_{[x}}^t\varepsilon_{yzt]}=0$, this condition can be solved if and only if $\tau=-\frac{\kappa}{2}$ and ${u_x}^y=\delta^y_x$. In conclusion, the quadratic constraint determines the embedding tensor \eqref{5+45theta} in terms of $J^m$ and $W^{mn}_p$ given by
\begin{eqnarray}\label{5+45JandW}
J^m&=&\kappa\delta^m_5,\qquad W^{ij}_5\ =\ W^{ij}_4\ =\ W^{xy}_z\ =\ W^{x5}_4\ =\ 0,\nonumber\\
W^{4y}_x&=&\lambda{w_x}^y,\ \quad W^{5x}_y\ =\ -\frac{\kappa}{2} \delta^x_y,\qquad W^{45}_x\ =\ v_x\, .
\end{eqnarray}
We note that $W^{45}_x$ written in terms of a three-dimensional vector $v_x$ in the last relation do not appear in the linear and quadratic constraints. Therefore, this vector is unconstrained.

With all these, non-vanishing gauge generators are given by
\begin{eqnarray}
X_{xy}&=&\sqrt{2}\varepsilon_{xyz}(\lambda{w_t}^z\boldsymbol{h}^{t5}-v_t\boldsymbol{h}^{tz}-2\kappa\boldsymbol{h}^{z4}),\label{5+45GG1}\\
X_{x5}&=&-\frac{\lambda}{\sqrt{2}}{w_x}^y\varepsilon_{yzt}\boldsymbol{h}^{zt},\label{5+45GG2}\\
X^x&=&4\kappa{\boldsymbol{\tau}^x}_5-2\lambda{w_y}^x{\boldsymbol{\tau}^y}_4,\label{5+45GG3}\\
X^4&=&2\lambda{w_x}^y{\boldsymbol{\tau}^x}_y+2v_x{\boldsymbol{\tau}^x}_5,\label{5+45GG4}\\
X^5&=&4\kappa({\boldsymbol{\tau}^4}_4+{\boldsymbol{\tau}^5}_5+\frac{2}{5}\boldsymbol{d})-2v_x{\boldsymbol{\tau}^x}_4\label{5+45GG5}
\end{eqnarray}
with the following commutation relations
\begin{eqnarray}
\left[X_{xy},X_{zt}\right]&=&\left[X_{xy},X_{z5}\right]\ =\ \left[X_{x5},X_{y5}\right]\ =\ \left[X^x,X^y\right]\ =\ \left[X^x,X_{y5}\right]\ =\ 0,\qquad\quad \nonumber\\
\left[X^x,X_{yz}\right]&=&16\kappa\delta^x_{[y}X_{z]5},\qquad\qquad\quad\left[X^4,X^x\right]\ =\ {(X^4)_y}^xX^y,\nonumber\\
\left[X^4,X_{x5}\right]&=&-{(X^4)_x}^yX_{y5},\qquad\quad\ \ \,\left[X^4,X_{xy}\right]\ =-{(X^4)_{xy}}^{zt}X_{zt}+8v_{[x}X_{y]5},\nonumber\\
\left[X^5,X^x\right]&=&-8\kappa X^x,\qquad\qquad\qquad\left[X^5,X_{x5}\right]\ =\ 0,\nonumber\\
\left[X^5,X_{xy}\right]&=&8\kappa X_{xy},\ \ \,\qquad\qquad\qquad\left[X^4,X^5\right]\ =\ 4v_xX^x.
\end{eqnarray}
\indent We find that the only possible compact subgroup is $SO(2)$ generated by the gauge generator $X^4$ with ${w_x}^y$ antisymmetric for any value of $v_x$. If the matrix ${w_x}^y$ is symmetric, the generator $X^4$ gives a non-compact $SO(1,1)$ group. For simplicity, we will set $v_x=0$ and restrict ourselves to the compact $SO(2)$ case since we are mainly interested in domain wall solutions preserving some symmetry. For definiteness, we choose the matrix ${w_x}^y$ to be
\begin{equation}
{w_x}^y=\begin{small}\begin{pmatrix} 0 & \sigma & 0 \\ -\sigma & 0 & 0 \\ 0 & 0 & 0 \end{pmatrix}\end{small}
\end{equation}
with $\sigma\in\mathbb{R}$. Together with $v_x=0$, the above gauge generators reduce to
\begin{eqnarray}
X_{12}&=-2\sqrt{2}\kappa\boldsymbol{h}^{34},\quad\qquad\qquad X_{\bar{x}3}&=-\sqrt{2}\varepsilon_{\bar{x}\bar{y}}(\lambda{w_{\bar{z}}}^{\bar{y}}\boldsymbol{h}^{\bar{z}5}-2\kappa\boldsymbol{h}^{\bar{y}4}),\nonumber\\
X_{\bar{x}5}&=-\sqrt{2}\lambda{w_{\bar{x}}}^{\bar{y}}\varepsilon_{\bar{y}\bar{z}}\boldsymbol{h}^{\bar{z}3},\quad\qquad X^{\bar{x}}&=\ 4\kappa{\boldsymbol{\tau}^{\bar{x}}}_5-2\lambda{w_{\bar{y}}}^{\bar{x}}{\boldsymbol{\tau}^{\bar{y}}}_4,\nonumber\\
X^3&=\ 4\kappa{\boldsymbol{\tau}^3}_5,\quad\qquad\qquad\qquad\  X^4&=\ 2\lambda\sigma({\boldsymbol{\tau}^1}_2-{\boldsymbol{\tau}^2}_1),\nonumber\\
X^5&=\ 4\kappa({\boldsymbol{\tau}^4}_4+{\boldsymbol{\tau}^5}_5+\frac{2}{5}\boldsymbol{d})\qquad\quad&\label{5+45xGG}
\end{eqnarray}
with $x=(\bar{x},3)$, $\bar{x}=1,2$. 

In vector representation, we can explicitly see that $X^4$ and $X^5$ are antisymmetric and symmetric
\begin{equation}
{(X^4)_m}^n=\begin{small}\begin{pmatrix} 0 & 4\lambda\sigma & 0 & 0 & 0 \\ -4\lambda\sigma & 0 & 0 & 0 & 0 \\ 0 & 0 & 0 & 0 & 0 \\ 0 & 0 & 0 & 0 & 0 \\ 0 & 0 & 0 & 0 & 0\end{pmatrix}\end{small},\qquad {(X^5)_m}^n=\begin{small}\begin{pmatrix} 0 & 0 & 0 & 0 & 0 \\ 0 & 0 & 0 & 0 & 0 \\ 0 & 0 & 0 & 0 & 0 \\ 0 & 0 & 0 & 8\kappa & 0 \\ 0 & 0 & 0 & 0 & 8\kappa \end{pmatrix}\end{small}
\end{equation}
generating $SO(2)$ and $SO(1,1)$ subgroups, respectively. The gauge generators satisfy the following commutation relations
\begin{eqnarray}
\left[X_{xy},X_{zt}\right]&=&\left[X_{xy},X_{\bar{z}5}\right]\ =\ \left[X_{\bar{x}5},X_{\bar{y}5}\right]\ =\ \left[X^x,X^y\right]\ =\ \left[X^x,X_{\bar{y}5}\right]\ =\ 0,\nonumber\\
\left[X^{\bar{x}},X_{12}\right]&=&16\kappa\delta^{\bar{x}}_{[1}X_{2]5},\qquad\qquad\quad\left[X^{\bar{x}},X_{\bar{y}3}\right]\ =\ 0,\nonumber\\
\left[X^3,X_{\bar{x}3}\right]&=&-8\kappa X_{\bar{x}5},\qquad\qquad\quad\ \ \left[X^3,X_{12}\right]\ =\ 0,\nonumber\\
\left[X^4,X^{\bar{x}}\right]&=&{(X^4)_{\bar{y}}}^{\bar{x}}X^{\bar{y}},\qquad\qquad\quad\,\left[X^4,X^3\right]\ =\ 0,\nonumber\\
\left[X^4,X_{\bar{x}3}\right]&=&-{(X^4)_{\bar{x}3}}^{\bar{y}3}X_{\bar{y}3},\qquad\quad\,\left[X^4,X_{12}\right]\ =\ 0,\nonumber\\
\left[X^4,X_{\bar{x}5}\right]&=&-{(X^4)_{\bar{x}}}^{\bar{y}}X_{\bar{y}5},\nonumber\\
\left[X^5,X^x\right]&=&-8\kappa X^x,\qquad\qquad\quad\ \ \, \left[X^5,X_{\bar{x}5}\right]\ =\ 0,\nonumber\\
\left[X^5,X_{xy}\right]&=&8\kappa X_{xy},\ \ \qquad\qquad\qquad\left[X^4,X^5\right]\ =\ 0\, .\label{45+5ComRels}
\end{eqnarray}
From these relations, we see that $(X^4,X_{\bar{x}5})$ and $(X^3,X^5,X_{12})$ form two commuting non-semisimple groups $ISO(2)$ and $ISO(1,1)$, respectively. The remaining generators $(X^{\bar{x}},X_{3\bar{x}})$ generate a four-dimensional translational group transforming non-trivially under $ISO(2)\times ISO(1,1)$. The gauge group is then given by
\begin{equation}\label{5+45gaugegroup}
G_0=\left(ISO(2)\times ISO(1,1)\right)\ltimes \mathbb{R}^4\, .
\end{equation}
\indent For a simpler case of $\lambda =0$, the non-vanishing components $W_y^{5x}$ still generate a non-trivial subgroup. In this case, we find $X^4=X_{\bar{x}5}=0$ giving rise to the following gauge group
\begin{equation}
SO(1,1)\ltimes\left(\mathbb{R}^3\times\mathbb{R}^3_{\boldsymbol{h}}\right).
\end{equation}
The three factors $SO(1,1)$, $\mathbb{R}^3$, and $\mathbb{R}^3_{\boldsymbol{h}}$ are respectively generated by $X^5$, $X_{xy}$, and $X^x$. We also note that another possibility of setting $\kappa=0$ is not possible since this choice leads to vanishing $J^m$. 

\subsection{Gaugings in $(\mathbf{10}+\mathbf{15})^{-1}$ representation}\label{10+15repSec}
For gaugings with the embedding tensor in $(\mathbf{10}+\mathbf{15})^{-1}$ representation, there are both electric and magnetic components $\theta^{Am}$ and $\theta^{A}_m$ given by
\begin{equation}\label{10+15theta}
\theta^{Am}\ =\ \frac{1}{3\sqrt{2}}\,\mathbb{T}^{A}_{np}\,\varepsilon^{mnpqr}Z_{qr}\qquad\text{ and }\qquad \theta^A_m\ = \ \mathbb{T}^{An}(Y_{nm}+Z_{nm}).
\end{equation}
We recall that $Y_{mn}=Y_{(mn)}$ and $Z_{mn}=Z_{[mn]}$ are symmetric and antisymmetric tensors corresponding to $\mathbf{15}^{-1}$ and $\mathbf{10}^{-1}$ representations, respectively. In terms of $Y_{mn}$ and $Z_{mn}$, the quadratic constraints \eqref{QC1} and \eqref{QC2} reduce to
\begin{equation}
\varepsilon^{npqrs}Y_{mq}Z_{rs}=0\qquad\text{and}\qquad \varepsilon^{mnpqr}Z_{np}Z_{qr}=0\, .\label{10+15QC}
\end{equation}
We can use the $SL(5)\subset GL(5)$ symmetry to bring $Y_{mn}$ to a diagonal form
\begin{equation}\label{diagYmn}
Y_{mn}\ =\ \text{diag}(\underbrace{1,..,1}_p,\underbrace{-1,..,-1}_q,\underbrace{0,..,0}_r)
\end{equation}
where $p+q+r=5$. We will split the $GL(5)$ index $m=(i,x)$ and $Y_{mn}$ into $Y_{ij}=\text{diag}(1,..,1,-1,..,-1)$ with $Y_{xy}=0$. Solving the first condition in \eqref{10+15QC} using \eqref{diagYmn}, we find only two possible solutions with both $Y_{mn}$ and $Z_{mn}$ non-vanishing for ranks of $Y_{mn}$ equal to $1$ and $2$.

\subsubsection{Rank$Y=2$}
With $Y_{ij}=\text{diag}(1,\pm1)$ for $i,j,...=1,2$ and $x,y,...=3,4,5$, only $Z_{12}=\kappa$ with $\kappa\in\mathbb{R}$ is allowed to be non-vanishing. This automatically solves the other condition in \eqref{10+15QC} and gives rise to the following gauge generators
\begin{eqnarray}
X_\ast&=&-\sqrt{2}\kappa\boldsymbol{h}^{12},\label{10+15Rank2Gen1}\\
X^x&=&-\frac{\sqrt{2}}{3}\kappa\varepsilon^{xyz}\boldsymbol{s}_{yz},\\
X_{ix}&=&2Y_{j[i}{\boldsymbol{\tau}^j}_{x]}-\frac{2}{3}\kappa\varepsilon_{j[i}{\boldsymbol{\tau}^j}_{x]},\\
X_{12}&=&2Y_{i[1}{\boldsymbol{\tau}^i}_{2]}+\frac{\kappa}{3}({\boldsymbol{\tau}^1}_{1}+{\boldsymbol{\tau}^2}_{2})+\frac{4}{5}\kappa\boldsymbol{d}\, .\label{10+15Rank2Gen4}
\end{eqnarray}
Commutation relations between these generators are given by
\begin{eqnarray}
\left[X_\ast,X^x\right]&=& \left[X_\ast,X_{ix}\right]\ = \ \left[X^x,X^y\right]\ = \ \left[X^x,X_{iy}\right]\ = \ \left[X_{ix},X_{jy}\right]\ = \ 0,\nonumber\\
\left[X_{12},X_\ast\right]&=&\frac{1}{2}{(X_{12})_m}^m X_\ast,\qquad\left[X_{12},X^x\right]\ =\ -2{(X_{12})_y}^x X^y,\nonumber\\
\left[X_{12},X_{ix}\right]&=&-{(X_{12})_i}^k X_{jx}-2{(X_{12})_x}^y X_{iy}\, .
\end{eqnarray}
\indent The generators $X_\ast$, $X^x$, and $X_{ix}$ generate three translational groups $\mathbb{R}_{\boldsymbol{h}}$, $\mathbb{R}^3_{\boldsymbol{s}}$, and $\mathbb{R}^6$ commuting with each other. For $Y_{ij}=\textrm{diag}(1,-1)$, the generator $X_{12}$ can only generate a non-compact $SO(1,1)$ group giving rise to the following gauge group
\begin{equation}
G_0=SO(1,1)\ltimes(\mathbb{R}^6\times\mathbb{R}^3_{\boldsymbol{s}}\times\mathbb{R}_{\boldsymbol{h}})\sim CSO(1,1,3)\ltimes(\mathbb{R}^3_{\boldsymbol{s}}\times\mathbb{R}_{\boldsymbol{h}}).\label{SO11_RankY2}
\end{equation}
For $Y_{ij}=\delta_{ij}$, $X_{12}$ can become compact $SO(2)$, non-compact $SO(1,1)$, or nilpotent generators depending on the values of $\kappa$. For particular values of $\kappa=\pm \sqrt{\frac{3}{5}}$, $X_{12}$ is nilpotent and will be denoted by $\mathcal{T}$ resulting in $\mc{T}\ltimes(\mathbb{R}^6\times\mathbb{R}^3_{\boldsymbol{s}}\times\mathbb{R}_{\boldsymbol{h}})$ gauge group. For $-\sqrt{\frac{3}{5}}<\kappa< \sqrt{\frac{3}{5}}$ and $\kappa<-\sqrt{\frac{3}{5}}$ or $\kappa>\sqrt{\frac{3}{5}}$, $X_{12}$ is respectively compact and non-compact. Accordingly, the corresponding gauge group is given by
\begin{equation}
G_0=SO(2)\ltimes(\mathbb{R}^6\times\mathbb{R}^3_{\boldsymbol{s}}\times\mathbb{R}_{\boldsymbol{h}})\sim CSO(2,0,3)\ltimes(\mathbb{R}^3_{\boldsymbol{s}}\times\mathbb{R}_{\boldsymbol{h}})\label{SO2_RankY2}
\end{equation}
or the gauge group given in \eqref{SO11_RankY2}.

\subsubsection{Rank$Y=1$}
In this case, we will choose only $Y_{11}$ non-vanishing with $Y_{11}=\kappa=\pm1$. All components $Z_{xy}$ with $x,y=2,3,4,5$ need to be zero in order to solve the quadratic constraint \eqref{10+15QC}. Therefore, only the remaining four components $Z_{1x}=z_x$ give rise to the following gauge generators
\begin{eqnarray}
X_\ast&=&-\sqrt{2}z_x\boldsymbol{h}^{1x},\qquad X^x\ =\ \frac{\sqrt{2}}{3}\varepsilon^{xyzt}z_y\boldsymbol{s}_{zt},\qquad
X_{xy}\ =-\frac{2}{3}z_{[x}{\boldsymbol{\tau}^1}_{y]},\nonumber\\
X_{1x}&=&\kappa{\boldsymbol{\tau}^1}_{x}+\frac{1}{3}(z_y{\boldsymbol{\tau}^y}_{x}+z_x{\boldsymbol{\tau}^1}_{1})+\frac{4}{5}z_x\boldsymbol{d}\label{10+15Rank1Gen4}
\end{eqnarray}
with the commutation relations
\begin{eqnarray}
\left[X_\ast,X^x\right]&=& \left[X_\ast,X_{xy}\right]\ = \ \left[X^x,X^y\right]\ = \ \left[X^x,X_{yz}\right]\ = \ \left[X_{xy},X_{zt}\right]\ = \ 0,\nonumber\\
\left[X_{1x},X_\ast\right]&=&\frac{1}{2}{(X_{1x})_m}^m X_\ast,\qquad\left[X_{1x},X^y\right]\ =\ -2{(X_{1x})_z}^y X^z,\nonumber\\
\left[X_{1x},X_{yz}\right]&=&\frac{1}{5}{(X_{1x})_{yz}}^{ta} X_{ta},\ \quad \left[X_{1x},X_{1y}\right]\ =\ -\frac{4}{3}z_{[x}X_{y]1}-2\kappa X_{xy}\, .
\end{eqnarray}
To proceed further, we will choose $z_2=\zeta$ and $z_3=z_4=z_5=0$. The above gauge generators simplify to
\begin{eqnarray}
X_\ast&=&-\sqrt{2}\zeta\boldsymbol{h}^{12},\qquad X^{\bar{x}}\ =-\frac{\sqrt{2}}{3}\zeta\varepsilon^{\bar{x}\bar{y}\bar{z}}\boldsymbol{s}_{\bar{y}\bar{z}},\nonumber\\
X_{2\bar{x}}&=&-\frac{1}{3}\zeta{\boldsymbol{\tau}^1}_{\bar{x}},\ \qquad X_{1\bar{x}}\ =\ \kappa{\boldsymbol{\tau}^1}_{\bar{x}}+\frac{1}{3}\zeta{\boldsymbol{\tau}^2}_{\bar{x}},\nonumber\\
X_{12}&=&\kappa{\boldsymbol{\tau}^1}_{2}+\frac{1}{3}\zeta({\boldsymbol{\tau}^2}_{2}+{\boldsymbol{\tau}^1}_{1})+\frac{4}{5}\zeta\boldsymbol{d}\label{10+15Rank1Gen4Ex}
\end{eqnarray}
with $\bar{x},\bar{y},...=3,4,5$. The corresponding commutation relations read
\begin{eqnarray}
\left[X_\ast,X^{\bar{x}}\right]&=& \left[X_\ast,X_{\bar{i}\bar{x}}\right]\ = \ \left[X^{\bar{x}},X^{\bar{y}}\right]\ = \ \left[X^{\bar{x}},X_{\bar{i}\bar{y}}\right]\ = \ \left[X_{\bar{i}\bar{x}},X_{\bar{j}\bar{y}}\right]\ = \ 0,\nonumber\\
\left[X_{12},X_\ast\right]&=&\frac{1}{2}{(X_{12})_m}^m X_\ast,\qquad\left[X_{12},X^{\bar{x}}\right]\ =\ -2{(X_{12})_{\bar{y}}}^{\bar{x}} X^{\bar{y}},\nonumber\\
\left[X_{12},X_{2\bar{x}}\right]&=&\frac{1}{5}{(X_{12})_{2\bar{x}}}^{2\bar{y}} X_{2\bar{y}},\qquad\left[X_{12},X_{1\bar{x}}\right]\ =\ \frac{2}{3}\zeta X_{1\bar{x}}-2\kappa X_{2\bar{x}}
\end{eqnarray}
with $\bar{i},\bar{j},...=1,2$. 

We also note that the generator $X_{12}$ does not commute with other generators. This generator generates an $SO(1,1)$ group while the remaining generators form three commuting translational groups. Accordingly, the resulting gauge group is given by
\begin{equation}
G_0=SO(1,1)\ltimes(\mathbb{R}^6\times\mathbb{R}^3_{\boldsymbol{s}}\times\mathbb{R}_{\boldsymbol{h}})
\end{equation}
in which the translational groups $\mathbb{R}^6$, $\mathbb{R}^3_{\boldsymbol{s}}$, and $\mathbb{R}_{\boldsymbol{h}}$ are respectively generated by $X_{\bar{i}\bar{x}}$, $X^{\bar{x}}$, and $X_\ast$. 

\subsection{Gaugings in $(\mathbf{10}+\overline{\mathbf{40}})^{-1}$ representation}\label{10+40repSec}
We now move to the next case with the embedding tensor in $\mathbf{10}^{-1}$ and $\overline{\mathbf{40}}^{-1}$ representations. These two components are labelled respectively by $Z_{mn}=Z_{[mn]}$ and $U^{mn,p}=U^{[mn],p}$ with $U^{[mn,p]}=0$. The embedding tensor solving the linear constraint \eqref{MainLC} takes the form
\begin{equation}\label{10+40theta}
\theta^{Am}\ =\ \mathbb{T}^{A}_{np}\left(U^{np,m} + \frac{1}{3\sqrt{2}}\varepsilon^{mnpqr}Z_{qr}\right)\qquad\text{ and }\qquad \theta^A_m\ = \ \mathbb{T}^{An}Z_{nm}\, .
\end{equation}
The quadratic constraint \eqref{QC1} imposes the following conditions
\begin{eqnarray}
U^{mn,p}Z_{pq}&=&0,\label{10+40QC1.1}\\
\varepsilon^{mnpqr}Z_{np}Z_{qr}&=&0\label{10+40QC1.2}
\end{eqnarray}
while the other condition in \eqref{QC2} gives
\begin{equation}
3\varepsilon_{mqrst}U^{qr,n}U^{st,p}+2\sqrt{2}Z_{mq}U^{qn,p}=0\, .\label{10+40QC2}
\end{equation}
The condition \eqref{10+40QC1.2} is similar to that given in the case of $\mathbf{10}^{-1}$ representation. Therefore, we will use the same ansatz for $Z_{mn}$ as given in \eqref{ZExform}. Moreover, as in section \ref{10Sec}, we will further fix each $GL(5)$ vector in the ansatz $Z_{mn}=u_{[m}v_{n]}$ in order to see the corresponding gauge group explicitly. However, instead of using $u_m=(\kappa_1,\kappa_2,0,0,0)$ and $v_m=(\lambda_1,\lambda_2,0,0,0)$, we equivalently set $Z_{12}=\kappa$ with other components vanishing. The condition \eqref{10+40QC1.2} is then automatically satisfied while the condition \eqref{10+40QC1.1} can be solved if and only if 
\begin{equation}
U^{12,m}=U^{ix,j}=U^{xy,i}=0\, .
\end{equation}
We have split the $GL(5)$ index as $m=(i,x)$ with $i,j,...=1,2$ and $x,y,...=3,4,5$. 

At this point, only $U^{ix,y}$ and $U^{xy,z}$ remain in terms of which the last condition in \eqref{10+40QC2} leads to
\begin{eqnarray}
\varepsilon_{ij}\varepsilon_{xta}U^{it,y}U^{ja,z}&=&0,\label{10+40QC2.1}\\
3\varepsilon_{ij}\varepsilon_{zta}(U^{jz,x}U^{ta,y}+U^{jz,y}U^{ta,x})+\sqrt{2}Z_{ij}U^{jx,y}&=&0\, .\label{10+40QC2.2}
\end{eqnarray}
These conditions can be readily solved by setting $U^{ix,y}=0$. This is very similar to gaugings from $(\mathbf{15}+\overline{\mathbf{40}})^{-1}$ representation with rank$Y=2$ studied in \cite{6D_DW_I}. Therefore, in order to solve the quadratic constraints for gaugings in $(\mathbf{10}+\overline{\mathbf{40}})^{-1}$ representation, we are left with only the following non-vanishing components
\begin{equation}\label{10+40Finaltensor}
Z_{ij}=\kappa\varepsilon_{ij}\qquad\text{and}\qquad U^{xy,z}=\frac{1}{2\sqrt{2}}\varepsilon^{xyt}{u_t}^z\, .
\end{equation}
We have parametrized the components $U^{xy,z}$ in terms of a traceless $3\times3$ matrix ${u_x}^y$ with ${u_x}^x=0$.

The corresponding gauge generators are given by
\begin{eqnarray}
X_\ast&=&-\sqrt{2}\kappa\boldsymbol{h}^{12},\label{10+45GenX1}\\
X^x&=&-\frac{1}{\sqrt{2}}\varepsilon^{xyz}{u_z}^t\boldsymbol{s}_{yt}-\frac{2\sqrt{2}}{3}\kappa\varepsilon^{xyz}\boldsymbol{s}_{yz},\\
X_{ix}&=&-\frac{1}{2}\varepsilon_{ij}{u_x}^y{\boldsymbol{\tau}^j}_{y}+\frac{1}{3}\kappa\varepsilon_{ij}{\boldsymbol{\tau}^j}_{x},\\
X_{12}&=&\frac{1}{2}{u_x}^y{\boldsymbol{\tau}^x}_{y}+\frac{\kappa}{3}({\boldsymbol{\tau}^2}_{2}+{\boldsymbol{\tau}^1}_{1})+\frac{4}{5}\kappa\boldsymbol{d} \label{10+45GenX12}
\end{eqnarray}
with the following commutation relations
\begin{eqnarray}
\left[X_\ast,X^x\right]&=&\left[X_\ast,X_{ix}\right]\ =\ \left[X^x,X^y\right]\ =\ \left[X^x,X_{iy}\right]\ =\ \left[X_{ix},X_{jy}\right]\ =\ 0,\nonumber\\
\left[X_{12},X_\ast\right]&=&\frac{1}{2}{(X_{12})_m}^mX_\ast,\quad\qquad\left[X_{12},X^x\right]={(X_{12})_y}^xX^y-4\kappa X^x,\nonumber\\
\left[X_{12},X_{ix}\right]&=&-2{(X_{12})_{ix}}^{jy}X_{jy}+4\kappa X_{ix}\, .\label{10+40ComRel}
\end{eqnarray}
The gauge generator $X_{12}$ given in \eqref{10+45GenX12} generates either $SO(1,1)$ or $SO(2)$ group for symmetric and antisymmetric ${u_x}^y$, respectively. 
The corresponding gauge group is given by 
\begin{equation}
G_0=SO(1,1)\ltimes(\mathbb{R}^6\times\mathbb{R}^3_{\boldsymbol{s}}\times\mathbb{R}_{\boldsymbol{h}})\sim CSO(1,1,3)\ltimes(\mathbb{R}^3_{\boldsymbol{s}}\times\mathbb{R}_{\boldsymbol{h}})\label{10+40GuageGroup}
\end{equation}
or
\begin{equation}
G_0=SO(2)\ltimes(\mathbb{R}^6\times\mathbb{R}^3_{\boldsymbol{s}}\times\mathbb{R}_{\boldsymbol{h}})\sim CSO(2,0,3)\ltimes(\mathbb{R}^3_{\boldsymbol{s}}\times\mathbb{R}_{\boldsymbol{h}})\label{10+40GuageGroupSO2}
\end{equation}
in which $X_{ix}$, $X^{x}$, and $X_\ast$ respectively generate the translational groups $\mathbb{R}^6$, $\mathbb{R}^3_{\boldsymbol{s}}$, and $\mathbb{R}_{\boldsymbol{h}}$.

\subsection{Gaugings in $(\mathbf{10}+\mathbf{15}+\overline{\mathbf{40}})^{-1}$ representation}\label{10+15+40repSec}
As a final example, we consider gaugings from $(\mathbf{10}+\mathbf{15}+\overline{\mathbf{40}})^{-1}$ representation in which the embedding tensor takes the form of
\begin{equation}\label{10+15+40theta}
\theta^{Am}\ =\ \mathbb{T}^{A}_{np}\left(U^{np,m} + \frac{1}{3\sqrt{2}}\varepsilon^{mnpqr}Z_{qr}\right)\qquad\text{ and }\qquad \theta^A_m\ = \ \mathbb{T}^{An}(Z_{nm}+Y_{mn}).
\end{equation}
The quadratic constraints in \eqref{QC1} and \eqref{QC2} give
\begin{eqnarray}
U^{mn,q}(3Z_{pq}+Y_{pq})+\frac{2\sqrt{2}}{3}\varepsilon^{mnqrs}Y_{pq}Z_{rs}&=&0,\\
U^{mn,q}(Z_{pq}+3Y_{pq})-\frac{\sqrt{2}}{3}\delta^{[m}_p\varepsilon^{n]qrst}Z_{qr}Z_{st}&=&0
\end{eqnarray}
together with
\begin{eqnarray}
U^{nq,m}Y_{pq}+\varepsilon^{mnqrs}Y_{pq}Z_{rs}-\frac{1}{2\sqrt{2}}U^{qr,m}U^{st,n}\varepsilon_{pqrst}&&\nonumber\\
+\frac{1}{3}\left[(2U^{mq,n}-U^{nq,m})Z_{pq}+4\delta^{(m}_pU^{n)qr}Z_{qr}\right]&&\nonumber\\
+\frac{1}{36\sqrt{2}}(\delta^m_p\varepsilon^{nqrst}-5\delta^n_p\varepsilon^{mqrst})Z_{qr}Z_{st}&=&0\, .
\end{eqnarray}
To solve these conditions, we repeat the same procedure as in the previous case by setting all $Z_{mn}$ components to zero except $Z_{12}=\kappa$ together with $Y_{mn}=\text{diag}(1,-1,0,0,0)$. With this choice together with the index splitting $m=(i,x)$ for $i,j,...=1,2$ and $x,y,...=3,4,5$, we can solve all the above conditions with all $U^{mn,p}$ components vanishing except
\begin{equation}
U^{xy,z}=\frac{1}{2\sqrt{2}}\varepsilon^{xyt}{u_t}^z\, .
\end{equation}
This is the same ansatz \eqref{10+40Finaltensor} as used in the previous case. 

These lead to the gauge generators 
\begin{eqnarray}
X_\ast&=&-\sqrt{2}\kappa\boldsymbol{h}^{12},\\
X^x&=&-\frac{1}{\sqrt{2}}\varepsilon^{xyz}{u_z}^t\boldsymbol{s}_{yt}-\frac{2\sqrt{2}}{3}\kappa\varepsilon^{xyz}\boldsymbol{s}_{yz},\\
X_{ix}&=&Y_{ji}{\boldsymbol{\tau}^j}_{x}-\frac{1}{2}\varepsilon_{ij}{u_x}^y{\boldsymbol{\tau}^j}_{y}+\frac{1}{3}\kappa\varepsilon_{ij}{\boldsymbol{\tau}^j}_{x},\\
X_{12}&=&2Y_{i[1}{\boldsymbol{\tau}^i}_{2]}+\frac{1}{2}{u_x}^y{\boldsymbol{\tau}^x}_{y}+\frac{\kappa}{3}({\boldsymbol{\tau}^2}_{2}+{\boldsymbol{\tau}^1}_{1})+\frac{4}{5}\kappa\boldsymbol{d}\label{10+15+40X12Gen}
\end{eqnarray}
with the same commutation relations as given in \eqref{10+40ComRel}. The corresponding gauge group is again given by \eqref{10+40GuageGroup} or \eqref{10+40GuageGroupSO2} depending on the values of $\kappa$ and the form of ${u_x}^y$ as in the previous case. It could be useful to look for more complicated solutions leading to new gauge groups. 

\section{Supersymmetric domain wall solutions}\label{DW_sol}
In this section, we will find supersymmetric domain wall solutions from six-dimensional gauged supergravities constructed from the embedding tensors given in the previous section. We begin with a general procedure of finding supersymmetric domain walls. The analysis has already been performed in \cite{6D_DW_I}, so we will only review relevant formulae and refer to \cite{6D_DW_I} for more detail. The metric ansatz for domain wall solutions take the general form of 
\begin{equation}\label{DWmetric}
ds_6^2=e^{2A(r)}\eta_{\bar{\mu} \bar{\nu}}dx^{\bar{\mu}} dx^{\bar{\nu}}+dr^2
\end{equation}
with $\bar{\mu},\bar{\nu}=0,1,\ldots,4$.
\\
\indent We then consider an explicit parametrization of the scalar fields in $SO(5,5)/SO(5)\times SO(5)$ coset space. These scalars correspond to $25$ non-compact generators of $SO(5,5)$ that decompose under $GL(5)$ as
\begin{equation}\label{15repscalarDEC}
25\ \rightarrow\ \underbrace{\mathbf{1}+\mathbf{14}}_{\hat{\boldsymbol{t}}^+_{a\dot{b}}}\,+\underbrace{\mathbf{10}}_{\boldsymbol{s}_{mn}}\, .
\end{equation}
The generators $\hat{\boldsymbol{t}}^+_{a\dot{b}}$ are non-compact generators of $GL(5)$ defined as
\begin{equation}
\hat{\boldsymbol{t}}^+_{a\dot{b}}=\frac{1}{2}\left({\mathbb{M}_{a}}^M{\mathbb{M}_{\dot{b}}}^N+{\mathbb{M}_{\dot{b}}}^M{\mathbb{M}_{a}}^N\right)\boldsymbol{t}_{MN}
\end{equation}
where ${\mathbb{M}_{\underline{A}}}^M=({\mathbb{M}_{a}}^M,{\mathbb{M}_{\dot{a}}}^M)$ is a transformation matrix whose explicit form is given by
\begin{equation}\label{offDiagTrans}
\mathbb{M}\ =\ \frac{1}{\sqrt{2}}\begin{pmatrix} \mathds{1}_5 & \mathds{1}_5 \\ \mathds{1}_5 & -\mathds{1}_5 \end{pmatrix}.
\end{equation}
This matrix is used to relate the $SO(5,5)$ metric in non-diagonal ($\eta_{MN}$) and diagonal ($\eta_{\ul{A}\ul{B}}$) forms as
\begin{equation}
\eta_{\ul{A}\ul{B}}={\mathbb{M}_{\underline{A}}}^M{\mathbb{M}_{\underline{B}}}^N\eta_{MN}
  \end{equation}  
with $\eta_{\ul{A}\ul{B}}=\textrm{diag}(1,1,1,1,1,-1,-1,-1,-1,-1)$.
\\  
\indent In accord with \eqref{15repscalarDEC}, we will denote all $25$ scalar fields as $\Phi^I=\{\varphi,\phi_1,\ldots,\phi_{14},\varsigma_1,\ldots,\varsigma_{10}\}$ with $I=1,\ldots,25$. The scalar $\varphi$ is the dilaton corresponding to the $SO(1,1)$ generator 
\begin{equation}\label{SO(1,1)Gen}
\boldsymbol{d}=\hat{\boldsymbol{t}}^+_{1\dot{1}}+\hat{\boldsymbol{t}}^+_{2\dot{2}}+\hat{\boldsymbol{t}}^+_{3\dot{3}}
+\hat{\boldsymbol{t}}^+_{4\dot{4}}+\hat{\boldsymbol{t}}^+_{5\dot{5}}\, .
\end{equation}
This $SO(1,1)\sim\mathbb{R}^+$  symmetry is identified with the $SO(1,1)$ factor of $GL(5)\sim SL(5)\times SO(1,1)$. The other fourteen linear combinations of generators $\hat{\boldsymbol{t}}^+_{a\dot{b}}$ correspond to the fourteen scalar fields $\{\phi_1,...,\phi_{14}\}$ in the $SL(5)/SO(5)$ coset. The remaining ten scalars $\{\varsigma_1,...,\varsigma_{10}\}$ correspond to the shift generators $\boldsymbol{s}_{mn}$ and will be called shift scalars. By this decomposition of the scalar fields, we can rewrite the kinetic terms of the scalar fields in \eqref{bosonic_L} and obtain the following bosonic Lagrangian
\begin{equation}
e^{-1}\mathcal{L}=\frac{1}{4}R-G_{IJ}\partial_\mu\Phi^I\partial^\mu\Phi^J-\mathbf{V}\label{Ex_bosonic_L}
\end{equation}
with $G_{IJ}$ being a symmetric scalar-dependent matrix.
\\
\indent In order to find supersymmetric domain wall solutions, we consider first-order Bogomol'nyi-Prasad-Sommerfield (BPS) equations derived from the supersymmetry transformations of fermionic fields. We begin with the variations of the gravitini from \eqref{1stSUSY} and \eqref{2ndSUSY} which give
\begin{eqnarray} 
\delta \psi_{+\bar{\mu}\alpha}&:&\quad A'\hat{\gamma}_{r}\epsilon_{+\alpha}+\frac{g}{2}\Omega_{\alpha\beta}T^{\beta\dot{\alpha}}\epsilon_{-\dot{\alpha}}=0,\label{eq1_GBPS}\\
\delta \psi_{-\bar{\mu}\dot{\alpha}}&:&\quad A'\hat{\gamma}_{r}\epsilon_{-\dot{\alpha}}-\frac{g}{2}\Omega_{\dot{\alpha}\dot{\beta}}T^{\alpha\dot{\beta}}\epsilon_{+\alpha}=0\, .\label{eq2_GBPS}
\end{eqnarray}
Throughout the paper, we use the notation $'$ to denote an $r$-derivative. Multiply equation \eqref{eq1_GBPS} by $A'\hat{\gamma}_{r}$ and use equation \eqref{eq2_GBPS} or vice-versa, we find the following consistency conditions
\begin{eqnarray}
{A'}^2{\delta_\alpha}^\beta &=&-\frac{g^2}{4}\Omega_{\alpha\gamma}T^{\gamma\dot{\alpha}}\Omega_{\dot{\alpha}\dot{\beta}}T^{\beta\dot{\beta}}=\mc{W}^2{\delta_\alpha}^\beta,\label{WarpBPSsq1}\\
{A'}^2{\delta_{\dot{\alpha}}}^{\dot{\beta}} &=&-\frac{g^2}{4}\Omega_{\dot{\alpha}\dot{\gamma}}T^{\alpha\dot{\gamma}}\Omega_{\alpha\beta}T^{\beta\dot{\beta}}
=\mc{W}^2{\delta_{\dot{\alpha}}}^{\dot{\beta}}\label{WarpBPSsq2}
\end{eqnarray}
in which we have introduced the ``superpotential'' $\mc{W}$. We then obtain the BPS equations for the warped factor
\begin{equation}
A'=\pm\mathcal{W}\, .\label{Ap_eq}
\end{equation}
With this result, equations \eqref{eq1_GBPS} and \eqref{eq2_GBPS} lead to the following projectors on the Killing spinors
\begin{equation}\label{DW_Proj}
\hat{\gamma}_r\epsilon_{+\alpha}=-\frac{g}{2}\Omega_{\alpha\beta}\frac{T^{\beta\dot{\beta}}}{A'}\epsilon_{-\dot{\beta}},\qquad
\hat{\gamma}_r\epsilon_{-\dot{\alpha}}=\frac{g}{2}\Omega_{\dot{\alpha}\dot{\beta}}\frac{T^{\alpha\dot{\beta}}}{A'}\epsilon_{+\alpha}.
\end{equation}
It should be noted that these projectors are not independent. The conditions $\delta \psi_{+r\alpha}=0$ and $\delta\psi_{-r\dot{\alpha}}=0$ will determine the Killing spinors as functions of the radial coordinate $r$ as usual. We will give the corresponding expressions in explicit solutions obtained in subsequent analyses.
\\
\indent Using the $\hat{\gamma}_r$ projectors in $\delta\chi_{+a\dot{\alpha}}=0$ and $\delta\chi_{-\dot{a}\alpha}=0$ equations, we eventually obtain the BPS equations for scalar fields of the form 
\begin{equation}\label{BPSGenEq}
{\Phi^I}'=\mp 2G^{IJ}\frac{\partial\mathcal{W}}{\partial\Phi^J}
\end{equation}
in which $G^{IJ}$ is the inverse of the scalar metric $G_{IJ}$. In addition, the scalar potential can also be written in terms of $\mc{W}$ as
\begin{equation}\label{PopularSP}
\mathbf{V}=2G^{IJ}\frac{\partial\mathcal{W}}{\partial\Phi^I}\frac{\partial\mathcal{W}}{\partial\Phi^J}-5\mathcal{W}^2\, .
\end{equation}
\indent We are now in a position to find explicit domain wall solutions. As mentioned before, the complexity of working with all $25$ scalars prohibits any traceable analysis. We will consider only solutions preserving a non-trivial residual symmetry. The solutions invariant under this symmetry are characterized by a smaller set of singlet scalars. Among the gauge groups classified in the previous section, only five gauge groups, $SO(2)\ltimes \mathbb{R}^8$, $SO(2)\ltimes \mathbb{R}^6$, $CSO(2,0,2)\ltimes \mathbb{R}^4$, $CSO(2,0,2)\ltimes \mathbb{R}^2$, and $CSO(2,0,1)\ltimes \mathbb{R}^4$, contain a compact $SO(2)$ subgroup that can be used to reduce the number of scalar fields and result in consistent sets of BPS equations. Accordingly, in the following, we will consider only domain wall solutions for these gauge groups. We should also remark here that for $\left(ISO(2)\times ISO(1,1)\right)\ltimes \mathbb{R}^4$ and $CSO(2,0,3)\ltimes \mathbb{R}^4$ gauge groups given in \eqref{5+45gaugegroup}, \eqref{SO2_RankY2}, and \eqref{10+40GuageGroupSO2}, we are not able to find consistent sets of BPS equations. The supersymmetry conditions from $(\delta\psi_{+\mu \alpha},\delta\psi_{-\mu \dot{\alpha}})$ and $(\delta \chi_{+a\dot{\alpha}},\delta \chi_{-\dot{a}\alpha})$ are not compatible with each other. Therefore, we argue that no supersymmetric domain walls with $SO(2)$ symmetry exist in these gauge groups.   

\subsection{Supersymmetric domain walls from $SO(2)\ltimes \mathbb{R}^8$ and $SO(2)\ltimes \mathbb{R}^6$ gauge groups}
We now consider supersymmetric domain walls from $SO(2)\ltimes \mathbb{R}^8$ and $SO(2)\ltimes \mathbb{R}^6$ gauge groups with the embedding tensor given in \eqref{theta_24_5} and the traceless matrix ${S_m}^n$ in \eqref{Schoices}. We first consider $SO(2)\ltimes \mathbb{R}^8$ gauge group. There are five $SO(2)$ singlet scalars corresponding to the following non-compact generators commuting with $X_\ast$
\begin{eqnarray}
\mathbf{Y}_1&=&\hat{\boldsymbol{t}}^+_{1\dot{1}}+\hat{\boldsymbol{t}}^+_{2\dot{2}}+\hat{\boldsymbol{t}}^+_{3\dot{3}}+\hat{\boldsymbol{t}}^+_{4\dot{4}}+\hat{\boldsymbol{t}}^+_{5\dot{5}},\label{24SO(2)non-com1}\\
\mathbf{Y}_2&=&\hat{\boldsymbol{t}}^+_{1\dot{1}}+\hat{\boldsymbol{t}}^+_{2\dot{2}}-2\,\hat{\boldsymbol{t}}^+_{3\dot{3}},\\
\mathbf{Y}_3&=&\hat{\boldsymbol{t}}^+_{4\dot{4}}+\hat{\boldsymbol{t}}^+_{5\dot{5}}-2\,\hat{\boldsymbol{t}}^+_{3\dot{3}},\\\mathbf{Y}_4&=&\boldsymbol{s}_{12},\\
\mathbf{Y}_5&=&\boldsymbol{s}_{45}\, .\label{24SO(2)non-com5}
\end{eqnarray}
It should be noted that the first non-compact generator corresponds to the $SO(1,1)$ factor defined in \eqref{SO(1,1)Gen}. This scalar is actually a singlet under the full compact subgroup $SO(5)\subset GL(5)$. The scalar field corresponding to this generator will be called the dilaton $\varphi$. Additionally, there are two $SL(5)/SO(5)$ scalars and two shift scalars invariant under the $SO(2)$ symmetry. These are associated with $\mathbf{Y}_{2,3}$ and $\mathbf{Y}_{4,5}$, respectively.

The coset representative can be written as
\begin{equation}\label{24SO(2)coset}
V=e^{\varphi\mathbf{Y}_1+\phi_1\mathbf{Y}_2+\phi_2\mathbf{Y}_3+\varsigma_1\mathbf{Y}_4+\varsigma_2\mathbf{Y}_5}\, .
\end{equation}
Using this coset representative, we find that the scalar potential vanishes identically, and only the dilaton appears in the T-tensors which take the form of
\begin{equation}\label{24T-tensorGen}
(T^a)^{\alpha\dot{\beta}}=\frac{1}{2\sqrt{2}}e^{5\varphi}{S_b}^a(\gamma^{b})^{\alpha\beta}\delta^{\dot{\beta}}_\beta,\qquad(T^{\dot{a}})^{\alpha\dot{\beta}}=\frac{1}{2\sqrt{2}}e^{5\varphi}{S_{\dot{b}}}^{\dot{a}}(\gamma^{\dot{b}})^{\dot{\alpha}\dot{\beta}}\delta_{\dot{\alpha}}^\alpha\, .
\end{equation}
In these equations, we have written ${S_b}^a={S_{\dot{b}}}^{\dot{a}}={S_m}^n$. With ${S_m}^n$ given in \eqref{Schoices}, the T-tensors become
\begin{equation}
T^{\alpha\dot{\beta}}=\frac{1}{\sqrt{2}}e^{5\varphi}\left(\lambda\, (\gamma^{45})^{\alpha\beta}-\kappa\, (\gamma^{12})^{\alpha\beta}\right)\delta^{\dot{\beta}}_\beta\qquad\qquad
\end{equation}
or explicitly 
\begin{equation}
\small
T^{\alpha\dot{\beta}}=\frac{1}{\sqrt{2}}e^{5\varphi}\begin{pmatrix} \lambda+\kappa & 0 & 0 & 0 \\ 0 & \lambda-\kappa & 0 & 0 \\ 0 & 0 & \lambda+\kappa & 0 \\ 0 & 0 & 0 & \lambda-\kappa\end{pmatrix}.
\end{equation}
There are two possible superpotentials
\begin{equation}\label{24SPot}
\mathcal{W}_1=\frac{g}{2\sqrt{2}}e^{5\varphi}(\kappa+\lambda)\qquad \text{and}\qquad
\mathcal{W}_2=\frac{g}{2\sqrt{2}}e^{5\varphi}(\kappa-\lambda)\, .
\end{equation}
Both of them give a valid superpotential in terms of which the scalar potential can be written as \eqref{PopularSP} using the matrix $G^{IJ}$ of the form 
\begin{equation}
G^{IJ}=\begin{small}\begin{pmatrix} \frac{1}{10} & 0 & 0 & \frac{2\varsigma_1}{5} & \frac{2\varsigma_2}{5}  \\
				0 & \frac{3}{20} & -\frac{1}{10} & \frac{3\varsigma_1}{5} & -\frac{2\varsigma_2}{5} \\
				0 &  -\frac{1}{10} & \frac{3}{20} & -\frac{2\varsigma_1}{5} & \frac{3\varsigma_2}{5} \\
				\frac{2\varsigma_1}{5} & \frac{3\varsigma_1}{5} & -\frac{2\varsigma_1}{5} & 1+4\varsigma_1^2 & 0 \\
				\frac{2\varsigma_2}{5} & -\frac{2\varsigma_2}{5} & \frac{3\varsigma_2}{5} & 0 & 1+4\varsigma_2^2 \\
		\end{pmatrix}\end{small}
\end{equation}
for $\Phi^I=\{\varphi,\phi_1,\phi_2,\varsigma_1,\varsigma_2\}$ with $I=1,2,3,4,5$. It should be noted that for $\lambda=0$, the two superpotentials are equal leading to half-supersymmetric domain walls. For $\lambda \neq 0$, one choice of the superpotentials gives rise to $\frac{1}{4}$-BPS domain walls since only the supersymmetry along the directions of this choice is unbroken.  

We now consider the cases of $\lambda\neq \pm \kappa$ and $\lambda \neq 0$ corresponding to $SO(2)\ltimes \mathbb{R}^8$ gauge group. In this case, choosing one of the two superpotentials corresponds to imposing an additional projector on the Killing spinors of the form
\begin{equation}\label{24Gam3Proj}
\gamma^3\epsilon_{\pm}=\epsilon_{\pm}\qquad \text{ or }\qquad \gamma^3\epsilon_{\pm}=-\epsilon_{\pm}
\end{equation}
for $\mc{W}=\mc{W}_1$ or $\mc{W}=\mc{W}_2$, respectively. Together with the $\hat{\gamma}_r$ projector in \eqref{DW_Proj}, the resulting solutions will preserve only eight supercharges or $\frac{1}{4}$ of the original supersymmetry. With all these, we obtain the following BPS equations
\begin{eqnarray}\label{24BPS}
A'&=&-\varphi'=\frac{g}{2\sqrt{2}}e^{5\varphi}(\kappa\pm\lambda),\\
\phi'_1&=&\phi'_2=0,\\
\varsigma'_1&=&4\varphi'\varsigma_1\ =-g\sqrt{2}e^{5\varphi}(\kappa\pm\lambda)\varsigma_1,\label{24anoshiftBPS1}\\ \varsigma'_2&=&4\varphi'\varsigma_2\ =-g\sqrt{2}e^{5\varphi}(\kappa\pm\lambda)\varsigma_2\, .\label{24anoshiftBPS2}
\end{eqnarray}
It should be noted that although the superpotential does not depend on $\varsigma_1$ and $\varsigma_2$, there are non-trivial BPS equations for these scalars due to the cross terms between these scalars and the dilaton in $G^{IJ}$.
\\
\indent The plus/minus signs in the BPS equations are correlated with the plus/minus sign of the $\gamma^3$ projector \eqref{24Gam3Proj}. The solution is given by
\begin{eqnarray}\label{24DW}
A&=&-\varphi=\frac{1}{5}\ln\left[\frac{5}{2\sqrt{2}}gr(\kappa\pm\lambda)-C\right],\\
\varsigma_{1,2}&=&e^{4\varphi}+\varsigma^{(0)}_{1,2}
\end{eqnarray}
for constants $C$ and $\varsigma^{(0)}_{1,2}$. The remaining two scalars $\phi_1$ and $\phi_2$ are constant. It turns out that in this case, all the composite connections $Q_\mu^{ab}$ and $Q_\mu^{\dot{a}\dot{b}}$ vanish. The BPS equations from $\delta \psi_{+r\alpha}$ and $\delta \psi_{-r\dot{\alpha}}$ then give the following Killing spinors
\begin{equation}\label{DWKilling}
\epsilon_{\pm}=e^{\frac{A(r)}{2}}\epsilon_{\pm}^{0}
\end{equation}
with the constant SMW spinors $\epsilon_{\pm}^{0}$ satisfying the projectors \eqref{DW_Proj} and \eqref{24Gam3Proj}. In this solution, an integration constant for $A$ has been neglected by rescaling the coordinates $x^{\bar\mu}$. Note that the integration constant $C$ and $\varsigma^{(0)}_{1,2}$ can also be removed by shifting the radial coordinate $r$ and scalars $\varsigma_{1,2}$, but we keep it here for generality. We also note that from the BPS equations, we can consistently set $\phi_1=\phi_2=\varsigma_1=\varsigma_2=0$. Indeed, it can be verified that redefining the shift scalars as $\varsigma_{1,2}\rightarrow \tilde{\varsigma}_{1,2} =e^{-4\varphi}\varsigma_{1,2}$ results in a set of BPS equations with $\tilde{\varsigma}'_{1,2}=0$. 

For $\lambda=\pm\kappa$ or $\lambda=0$, the translational group $\mathbb{R}^{8}_{\boldsymbol{s}}$ reduces to $\mathbb{R}^{6}_{\boldsymbol{s}}$. In the former case, $X_{14}=\pm X_{25}$ and $X_{15}=\mp X_{24}$ while in the latter $X_{34}$ and $X_{35}$ vanish. In these two cases, there are more $SO(2)$ singlet scalars. For $\lambda=\pm\kappa$, there are four additional singlet scalars corresponding to non-compact generators
\begin{eqnarray}
\mathbf{Y}_6&=&\hat{\boldsymbol{t}}^+_{1\dot{4}}\pm\hat{\boldsymbol{t}}^+_{2\dot{5}},\qquad\,
\mathbf{Y}_7\ =\ \hat{\boldsymbol{t}}^+_{1\dot{5}}\mp\hat{\boldsymbol{t}}^+_{2\dot{4}},\nonumber \\
\mathbf{Y}_8&=&\boldsymbol{s}_{14}\pm\boldsymbol{s}_{15},\qquad
\mathbf{Y}_9\ =\ \boldsymbol{s}_{15}\mp\boldsymbol{s}_{24}\, .
\end{eqnarray}
The upper/lower signs in these generators are related to the upper/lower sign in  $\lambda=\pm\kappa$. For $\lambda=0$, we find six additional singlets given by
\begin{eqnarray}
\widetilde{\mathbf{Y}}_6&=&\hat{\boldsymbol{t}}^+_{4\dot{4}}-\hat{\boldsymbol{t}}^+_{5\dot{5}},\qquad
\widetilde{\mathbf{Y}}_7\ =\ \hat{\boldsymbol{t}}^+_{3\dot{4}},\qquad
\widetilde{\mathbf{Y}}_8\ =\ \hat{\boldsymbol{t}}^+_{3\dot{5}},\nonumber \\
\widetilde{\mathbf{Y}}_9&=&\hat{\boldsymbol{t}}^+_{4\dot{5}},\qquad\qquad
\widetilde{\mathbf{Y}}_{10}\ =\ \boldsymbol{s}_{34},\quad\ \,
\widetilde{\mathbf{Y}}_{11}\ =\ \boldsymbol{s}_{35}\, . \label{24addgenkap0}
\end{eqnarray}
It turns out that even with all these extra $SO(2)$ singlet scalars, the scalar potential still vanishes identically. Moreover, the T-tensors depend only on the dilaton and take the same form as given in \eqref{24T-tensorGen}. The resulting BPS equations for scalars from $SL(5)/SO(5)$ associated with $\mathbf{Y}_{6,7}$ and $\widetilde{\mathbf{Y}}_{6,7,8,9}$ give constant scalars. Similar to the previous case, the shift scalars associated with $\mathbf{Y}_{8,9}$ and $\widetilde{\mathbf{Y}}_{10,11}$ can be redefined such that the corresponding BPS equations give constant scalars. All these constant scalars can in turn be consistently set to zero. Therefore, the solutions for these two special cases are given by the above solution with the substitution $\lambda =\pm \kappa$ and $\lambda=0$. However, as mentioned before, solutions with $\lambda=0$ is $\frac{1}{2}$-supersymmetric since in this case the two superpotentials $\mc{W}_1$ and $\mc{W}_2$ are equal. On the other hand, solutions with $\lambda =\pm \kappa$ preserve $\frac{1}{4}$ of the original supersymmetry as in the more general case of $\lambda \neq\pm \kappa$. Each choice of $\lambda =\pm \kappa$ makes one of the superpotentials vanish rendering the supersymmetry along the directions of this superpotential broken.  

\subsection{Supersymmetric domain walls from $CSO(2,0,2)\ltimes \mathbb{R}^4$, $CSO(2,0,2)\ltimes \mathbb{R}^2$, and $CSO(2,0,1)\ltimes \mathbb{R}^4$ gauge groups}
We now move to another class of solutions in $CSO(2,0,2)\ltimes \mathbb{R}^4$, $CSO(2,0,2)\ltimes \mathbb{R}^2$, and $CSO(2,0,1)\ltimes \mathbb{R}^4$ gauge groups obtained from the embedding tensor in $\overline{\mathbf{45}}^{+3}$ with the traceless matrix ${u_i}^j$ given in \eqref{uchoices}. Since in this case, the three possibilities with $\lambda\neq \pm \kappa$, $\lambda=\pm \kappa$, and $\lambda=0$ are different in many aspects, we will separately consider these three cases.

\subsubsection{$SO(2)$ symmetric domain walls from $CSO(2,0,2)\ltimes \mathbb{R}^4$ gauge group}
With $\lambda\neq\pm\kappa$, the gauge group is given by $CSO(2,0,2)\ltimes \mathbb{R}^4$ as in \eqref{45RepSO2GG}. For convenience, we note the explicit form of the matrix ${u_i}^j$ of the gauge group \eqref{45RepSO2GG}  
\begin{equation}\label{SO(2)uchoices}
{u_i}^j=\begin{small}\begin{pmatrix} 0 & \kappa & 0 & 0\\ -\kappa & 0 & 0 & 0\\  0 & 0 & 0 & -\lambda\\ 0 & 0 & \lambda & 0 \end{pmatrix}\end{small}
\end{equation}
where $\kappa$ and $\lambda$ are non-vanishing and $\lambda\neq\pm\kappa$. There are five $SO(2)$ singlet scalars containing the dilaton, two scalars from $SL(5)/SO(5)$, and two shift scalars corresponding to the following non-compact generators 
\begin{eqnarray}
\overline{\mathbf{Y}}_1&=&\hat{\boldsymbol{t}}^+_{1\dot{1}}+\hat{\boldsymbol{t}}^+_{2\dot{2}}+\hat{\boldsymbol{t}}^+_{3\dot{3}}+\hat{\boldsymbol{t}}^+_{4\dot{4}}+\hat{\boldsymbol{t}}^+_{5\dot{5}},\label{45SO(2)non-com1}\\
\overline{\mathbf{Y}}_2&=&\hat{\boldsymbol{t}}^+_{1\dot{1}}+\hat{\boldsymbol{t}}^+_{2\dot{2}}+\hat{\boldsymbol{t}}^+_{3\dot{3}}+\hat{\boldsymbol{t}}^+_{4\dot{4}}-4\hat{\boldsymbol{t}}^+_{5\dot{5}},\\
\overline{\mathbf{Y}}_3&=&\hat{\boldsymbol{t}}^+_{1\dot{1}}+\hat{\boldsymbol{t}}^+_{2\dot{2}}-\hat{\boldsymbol{t}}^+_{3\dot{3}}-\hat{\boldsymbol{t}}^+_{4\dot{4}},\\
\overline{\mathbf{Y}}_4&=&\boldsymbol{s}_{12},\\
\overline{\mathbf{Y}}_5&=&\boldsymbol{s}_{34}\, .\label{45SO(2)non-com5}
\end{eqnarray}
The first non-compact generator corresponds to the dilaton $\varphi$ as in the previous section. The scalar field corresponding to $\overline{\mathbf{Y}}_2$ is another dilaton coming from further decomposition of $SL(5)\rightarrow SO(1,1)\times SL(4)$ and will be denoted by $\phi_0$. With the split of $GL(5)$ index $m=(i,5)$ for $i=1,...,4$, the $SL(5)/SO(5)$ coset also decomposes into $SO(1,1)\times SL(4)/SO(4)$ with the non-compact generator $\overline{\mathbf{Y}}_3$ corresponding to an $SO(2)$ singlet from the $SL(4)/SO(4)$ coset.

With the coset representative given by
\begin{equation}\label{45SO(2)coset}
V=e^{\varphi\overline{\mathbf{Y}}_1+\phi_0\overline{\mathbf{Y}}_2+\phi\overline{\mathbf{Y}}_3
+\varsigma_1\overline{\mathbf{Y}}_4+\varsigma_2\overline{\mathbf{Y}}_5},
\end{equation}
the scalar potential vanishes identically as in the previous section. The T-tensor is more complicated due to the dependence on the shift scalars with the following form
\begin{eqnarray}
T^{\alpha\dot{\beta}}=e^{-3\varphi-8\phi_0}\left[2(\lambda\varsigma_1-\kappa\varsigma_2)\, \Omega^{\alpha\beta}+2(\kappa\varsigma_1-\lambda\varsigma_2)\, (\gamma^{5})^{\alpha\beta}\phantom{\frac{1}{2}}\right.\nonumber\\ \left.\phantom{\frac{1}{2}}+(\lambda+4\kappa\varsigma_1\varsigma_2)\, (\gamma^{12})^{\alpha\beta}+(\kappa+4\lambda\varsigma_1\varsigma_2)\, (\gamma^{34})^{\alpha\beta}\right]\delta^{\dot{\beta}}_\beta\, .\label{45T-tensor}
\end{eqnarray}
This gives rise to
\begin{eqnarray}
\Omega_{\alpha\gamma}T^{\gamma\dot{\alpha}}\Omega_{\dot{\alpha}\dot{\beta}}T^{\beta\dot{\beta}}&=&-e^{-6\varphi-16\phi_0}(1+4\varsigma_1^2)(1+4\varsigma_2^2)\left[(\kappa^2+\lambda^2){\delta_\alpha}^\beta+2\kappa\lambda{(\gamma^5)_\alpha}^\beta\right],\nonumber\\&&\label{45PreSPot1}\\
\Omega_{\dot{\alpha}\dot{\gamma}}T^{\alpha\dot{\gamma}}\Omega_{\alpha\beta}T^{\beta\dot{\beta}}&=&-e^{-6\varphi-16\phi_0}(1+4\varsigma_1^2)(1+4\varsigma_2^2)\left[(\kappa^2+\lambda^2){\delta_{\dot{\alpha}}}^{\dot{\beta}}+2\kappa\lambda{(\gamma^{\dot{5}})_{\dot{\alpha}}}^{\dot{\beta}}\right].\nonumber\\\label{45PreSPot2}
\end{eqnarray}
\indent In this case, the two shift scalars $\varsigma_1$ and $\varsigma_2$ cannot be removed by a redefinition of scalar fields, and the composite connections $Q_\mu^{ab}$ and $Q_\mu^{\dot{a}\dot{b}}$ are non-vanishing. Therefore, as in \cite{6D_DW_I}, we need to modify the ansatz for the Killing spinors to 
\begin{equation}\label{fisrtModKilling}
\epsilon_+=e^{\frac{A(r)}{2}+B(r)\gamma_{12}}\epsilon_{+}^{0}\qquad\text{and}\qquad\epsilon_-=e^{\frac{A(r)}{2}-B(r)\gamma_{\dot{1}\dot{2}}}\epsilon_{-}^{0}
\end{equation}
in which $B(r)$ is an $r$-dependent function. We now impose the $\gamma^5$ projector 
\begin{equation}\label{45Gam5Proj}
\gamma^5\epsilon_{\pm}=\epsilon_{\pm}\qquad \text{ or }\qquad \gamma^5\epsilon_{\pm}=-\epsilon_{\pm}
\end{equation}
which implies $\gamma_{12}\epsilon_+=\mp\gamma_{34}\epsilon_+$ and $\gamma_{\dot{1}\dot{2}}\epsilon_-=\mp\gamma_{\dot{3}\dot{4}}\epsilon_-$. The latter two conditions are sufficient to solve the conditions $\delta \psi_{+r\alpha}=0$ and $\delta\psi_{-r\dot{\alpha}}=0$ with $\epsilon_{\pm}^{0}$ being constant SMW spinors making $\epsilon_\pm$ satisfy the projectors \eqref{DW_Proj} and \eqref{45Gam5Proj}.
\\
\indent With the projector \eqref{45Gam5Proj}, the two conditions in \eqref{45PreSPot1} and \eqref{45PreSPot2} give
\begin{eqnarray}
\Omega_{\alpha\gamma}T^{\gamma\dot{\alpha}}\Omega_{\dot{\alpha}\dot{\beta}}T^{\beta\dot{\beta}}\epsilon_{+\beta}&=&-e^{-6\varphi-16\phi_0}(1+4\varsigma_1^2)(1+4\varsigma_2^2)(\kappa\pm\lambda)^2\epsilon_{+\alpha},\\
\Omega_{\dot{\alpha}\dot{\gamma}}T^{\alpha\dot{\gamma}}\Omega_{\alpha\beta}T^{\beta\dot{\beta}}\epsilon_{-\dot{\beta}}&=&-e^{-6\varphi-16\phi_0}(1+4\varsigma_1^2)(1+4\varsigma_2^2)(\kappa\pm\lambda)^2\epsilon_{-\dot{\alpha}}
\end{eqnarray}
in which the plus/minus signs correspond to the sign chosen in the $\gamma^5$ projector. Therefore, the superpotential is then given by
\begin{equation}
\mc{W}=\frac{g}{2}e^{-3\varphi-8\phi_0}(\kappa\pm\lambda)\sqrt{1+4\varsigma_1^2}\sqrt{1+4\varsigma_2^2}\, .\label{45SPot}
\end{equation}
With all these, we find the following BPS equations
\begin{eqnarray}
A'&=&\frac{g}{2}e^{-3\varphi-8\phi_0}(\kappa\pm\lambda)\sqrt{1+4\varsigma_1^2}\sqrt{1+4\varsigma_2^2},\label{45BPS1}\\
\varphi'&=&\frac{ge^{-3\varphi-8\phi_0}(\kappa\pm\lambda)\left[3-4(\varsigma_1^2+\varsigma_2^2)-80\varsigma_1^2\varsigma_2^2\right]}{10\sqrt{1+4\varsigma_1^2}\sqrt{1+4\varsigma_2^2}},\\
\phi_0'&=&\frac{ge^{-3\varphi-8\phi_0}(\kappa\pm\lambda)\left[1+4(\varsigma_1^2+\varsigma_2^2)\right]}{5\sqrt{1+4\varsigma_1^2}\sqrt{1+4\varsigma_2^2}},\\
\phi'&=&\frac{2ge^{-3\varphi-8\phi_0}(\kappa\pm\lambda)\left(\varsigma_2^2-\varsigma_1^2\right)}{\sqrt{1+4\varsigma_1^2}\sqrt{1+4\varsigma_2^2}},\label{45BPS4}\\
\varsigma'_1&=&-2ge^{-3\varphi-8\phi_0}(\kappa\pm\lambda)\varsigma_1\sqrt{1+4\varsigma_1^2}\sqrt{1+4\varsigma_2^2},\label{45BPS5}\\
\varsigma'_2&=&-2ge^{-3\varphi-8\phi_0}(\kappa\pm\lambda)\varsigma_2\sqrt{1+4\varsigma_1^2}\sqrt{1+4\varsigma_2^2},\label{45BPS6}\\
B'&=&\frac{2ge^{-3\varphi-8\phi_0}(\kappa\pm\lambda)(\varsigma_1-\varsigma_2-4\varsigma^2_1\varsigma_2+4\varsigma_1\varsigma^2_2)}{\sqrt{1+4\varsigma_1^2}\sqrt{1+4\varsigma_2^2}}\, .\label{45B'}
\end{eqnarray}
The domain wall solution to these equations can be written in terms of the shift scalar $\varsigma_1$ by
\begin{eqnarray}
A&=&-\frac{1}{4}\ln\varsigma_1,\qquad\qquad
\varsigma_2\ =\ C_1\varsigma_1,\label{45DW1}\\
\phi&=&C_2+\frac{1}{8}\ln\left(1+4\varsigma_1^2\right)-\frac{1}{8}\ln\left(1+4C^2_1\varsigma_1^2\right),\\
\phi_0&=&C_3-\frac{1}{10}\ln\varsigma_1+\frac{1}{40}\ln\left(1+4\varsigma_1^2\right)+\frac{1}{40}\ln\left(1+4C^2_1\varsigma_1^2\right),\\
\varphi&=&C_4-\frac{3}{20}\ln\varsigma_1+\frac{1}{10}\ln\left(1+4\varsigma_1^2\right)+\frac{1}{10}\ln\left(1+4C^2_1\varsigma_1^2\right),\label{45DW5}\\
B&=& C_5-\frac{1}{2}\tan^{-1}(2\varsigma_1)+\frac{1}{2}\tan^{-1}(2C_1\varsigma_1)\label{45DW6}
\end{eqnarray}
in which $C_i$, $i=1,...,5$, are integration constants. Using the new radial coordinate $\rho$ defined by $\frac{d\rho}{dr}=e^{-3\varphi-8\phi_0}$, we find the solution for $\varsigma_1$ given by
\begin{equation}
\varsigma_1=\frac{\text{sech}[2g(\kappa\pm\lambda)\rho-C]}{2\sqrt{C_1^2\tanh^2[2g(\kappa\pm\lambda)\rho-C]-1}}.\label{45DW7}
\end{equation}
\indent It is also interesting to consider a domain wall solution with vanishing shift scalars. From the above solution, we find a constant $B$ function giving rise to the same Killing spinors \eqref{DWKilling} up to a constant phase. For $\varsigma_1=\varsigma_2=0$, the BPS equations \eqref{45BPS1} to \eqref{45BPS6} reduce to 
\begin{eqnarray}
A'&=&\frac{g}{2}(\kappa\pm\lambda) e^{-3\varphi-8\phi_0},\\  
\varphi'&=&\frac{3}{10}g(\kappa\pm\lambda) e^{-3\varphi-8\phi_0},\\
\phi'_0&=&\frac{g}{5}(\kappa\pm\lambda) e^{-3\varphi-8\phi_0}
\end{eqnarray}
together with $\phi'=\varsigma'_1=\varsigma'_2=0$. Setting $\phi=\varsigma_1=\varsigma_2=0$, we readily find the solution
\begin{eqnarray}
A&=&\frac{5}{2}\phi_0,\qquad \varphi\ =\ \frac{1}{3}\ln\left[\frac{5}{2}[g(\kappa\pm\lambda) r+C]\right]-\frac{8}{3}\phi_0,\nonumber\\ \phi_0&=&C_0+\frac{2}{25}\ln\left[g(\kappa\pm\lambda) r+C\right]\label{45SpeDW}.
\end{eqnarray}
Since the solutions are subject to two projectors \eqref{DW_Proj} and \eqref{45Gam5Proj}, all the domain walls, with and without the shift scalars, are $\frac{1}{4}$-supersymmetric. 

\subsubsection{$SO(2)$ symmetric domain walls from $CSO(2,0,2)\ltimes \mathbb{R}^2$ gauge group}
For $\lambda=\pm\kappa$, we find that the gauge generators $X_{13}=\pm X_{24}$ and $X_{14}=\mp X_{23}$. This reduces the translational group $\mathbb{R}^4_{\boldsymbol{h}}$ to $\mathbb{R}^2_{\boldsymbol{h}}$ resulting in the gauge group of the form
\begin{equation}
G_0=SO(2)\ltimes\left(\mathbb{R}^4\times\mathbb{R}^2_{\boldsymbol{h}}\right)\sim CSO(2,0,2)\ltimes\mathbb{R}^2_{\boldsymbol{h}}.
\end{equation}
Apart from the five singlet scalars corresponding to non-compact generators in \eqref{45SO(2)non-com1} to \eqref{45SO(2)non-com5}, there are additional four scalars invariant under the $SO(2)$ subgroup generated by $X^5$ with $\lambda=\pm\kappa$. These extra scalars correspond to the following non-compact generators
\begin{eqnarray}
\overline{\mathbf{Y}}_6&=&\hat{\boldsymbol{t}}^+_{1\dot{3}}\mp\hat{\boldsymbol{t}}^+_{2\dot{4}},\qquad \overline{\mathbf{Y}}_7\ =\ \hat{\boldsymbol{t}}^+_{1\dot{4}}\pm\hat{\boldsymbol{t}}^+_{2\dot{3}},\nonumber \\
\overline{\mathbf{Y}}_8&=&\boldsymbol{s}_{13}\mp\boldsymbol{s}_{24},\, \qquad\overline{\mathbf{Y}}_9\ =\ \boldsymbol{s}_{14}\pm\boldsymbol{s}_{23}\, .
\end{eqnarray}
Using the coset representative 
\begin{equation}\label{45SO(2)cosetlam=kap}
V=e^{\varphi\overline{\mathbf{Y}}_1+\phi_0\overline{\mathbf{Y}}_2+\phi_2\overline{\mathbf{Y}}_6+\phi_3\overline{\mathbf{Y}}_7+\phi\overline{\mathbf{Y}}_3+\varsigma_1\overline{\mathbf{Y}}_4+\varsigma_2\overline{\mathbf{Y}}_5+\varsigma_3\overline{\mathbf{Y}}_8+\varsigma_4\overline{\mathbf{Y}}_9},
\end{equation}
we again find that the scalar potential vanishes identically. Using the $\gamma^5$ projector given in \eqref{45Gam5Proj}, we find that the superpotential is given by
\begin{equation}
\mc{W}=ge^{-3\varphi-8\phi_0}\kappa\sqrt{4\varsigma^2_1+4\varsigma^2_2+16\varsigma^2_1\varsigma^2_2
+32\varsigma_1\varsigma_2(\varsigma^2_3+\varsigma^2_4)+(1+4\varsigma^2_3+4\varsigma^2_4)^2}\, .\label{45SPotlam=kap}
\end{equation}
Consistency of the BPS equations requires either $\varsigma_3=\varsigma_4=0$ or $\varsigma_2=\varsigma_1$. The former gives the same solution as in the previous case with ($\kappa\pm\lambda$) replaced by $2\kappa$, so we will not consider this choice any further. 

For the other choice of $\varsigma_2=\varsigma_1$, the superpotential \eqref{45SPotlam=kap} becomes
\begin{equation}
\mc{W}=ge^{-3\varphi-8\phi_0}\kappa\left(1+4\varsigma^2_1+4\varsigma^2_3+4\varsigma^2_4\right).\label{new45SPotlam=kap}
\end{equation}
In this case, the composite connections $Q_\mu^{ab}$ and $Q_\mu^{\dot{a}\dot{b}}$ vanish. We can then use the ansatz for the Killing spinors as given in \eqref{DWKilling} and end up with the following BPS equations
\begin{eqnarray}
A'&=&ge^{-3\varphi-8\phi_0}\kappa\left(1+4\varsigma^2_1+4\varsigma^2_3+4\varsigma^2_4\right),\label{45newBPS1}\\
\varphi'&=&\frac{g}{5}e^{-3\varphi-8\phi_0}\kappa(3-20\varsigma^2_1-20\varsigma^2_3-20\varsigma^2_4),\label{45newBPS2}\\
\phi_0'&=&\frac{2g}{5}e^{-3\varphi-8\phi_0}\kappa,\label{45newBPS3}\\
\varsigma'_1&=&-4ge^{-3\varphi-8\phi_0}\kappa\varsigma_1\left(1+4\varsigma^2_1+4\varsigma^2_3+4\varsigma^2_4\right),\label{45BPS4}\\
\varsigma'_3&=&-4ge^{-3\varphi-8\phi_0}\kappa\varsigma_3\left(1+4\varsigma^2_1+4\varsigma^2_3+4\varsigma^2_4\right),\label{45BPS5}\\
\varsigma'_4&=&-4ge^{-3\varphi-8\phi_0}\kappa\varsigma_4\left(1+4\varsigma^2_1+4\varsigma^2_3+4\varsigma^2_4\right)
\end{eqnarray}
together with $\phi'=\phi'_2=\phi'_3=0$. Solving these equations gives a domain wall solution 
\begin{eqnarray}
A&=&-\frac{1}{4}\ln\varsigma_1,\qquad \varsigma_3\ =\ C_1\varsigma_1,\qquad \varsigma_4\ =\ C_2\varsigma_1,\nonumber\\
\varphi&=&C_3-\frac{3}{20}\ln{\varsigma_1}+\frac{1}{5}\ln\left[1+4\varsigma^2_1(1+C^2_1+C^2_2)\right],\nonumber\\
\phi_0&=&C_4-\frac{1}{10}\ln{\varsigma_1}+\frac{1}{20}\ln\left[1+4\varsigma^2_1(1+C^2_1+C^2_2)\right],\nonumber\\
\varsigma_1&=&\frac{1}{\sqrt{e^{8g\kappa\rho-C_5}-4(1+C^2_1+C^2_2)}}
\end{eqnarray}
in which $\rho$ is a new radial coordinate defined by the relation $\frac{d\rho}{dr}=e^{-3\varphi-8\phi_0}$. In this solution, we also consistently set the constant scalars $\phi$, $\phi_2$, and $\phi_3$ to zero. 

\subsubsection{$SO(2)$ symmetric domain walls from $CSO(2,0,1)\ltimes \mathbb{R}^4$ gauge group}
For $\lambda=0$, we find $X^3=X^4=0$ reducing $\mathbb{R}^4$ to $\mathbb{R}^2$ in the gauge group \eqref{45RepSO2GG}. The resulting gauge group then takes the form of
\begin{equation}
G_0=SO(2)\ltimes\left(\mathbb{R}^2\times\mathbb{R}^4_{\boldsymbol{h}}\right)\sim CSO(2,0,1)\ltimes\mathbb{R}^4_{\boldsymbol{h}}\, .
\end{equation}
In this case, the $\gamma^5$ projector \eqref{45Gam5Proj} is not needed, so unlike the previous two cases, the solutions will be $\frac{1}{2}$-supersymmetric. There are additional scalars invariant under the $SO(2)$ subgroup generated by $X^5$ with $\lambda=0$. These are given by two scalars parametrizing the $SL(4)/SO(4)$ coset, two nilpotent, and two shift scalars corresponding to the following non-compact generators
\begin{eqnarray}
\widehat{\mathbf{Y}}_6&=&\hat{\boldsymbol{t}}^+_{3\dot{3}}-\hat{\boldsymbol{t}}^+_{4\dot{4}},\qquad \widehat{\mathbf{Y}}_7\ =\ \hat{\boldsymbol{t}}^+_{3\dot{4}},\qquad
\widehat{\mathbf{Y}}_8\ =\ \hat{\boldsymbol{t}}^+_{3\dot{5}},\nonumber \\ 
\widehat{\mathbf{Y}}_9&=&\hat{\boldsymbol{t}}^+_{4\dot{5}},\qquad\qquad
\widehat{\mathbf{Y}}_{10}\ =\ \boldsymbol{s}_{35},\ \,\quad\widehat{\mathbf{Y}}_{11}\ =\ \boldsymbol{s}_{45}\, .
\end{eqnarray}
Following \cite{6D_DW_I}, we will call the four scalars associated with $\hat{\boldsymbol{t}}^+_{i\dot{5}}$, for $i=1,2,3,4$, nilpotent scalars. It turns out that the two nilpotent scalars associated with $\hat{\boldsymbol{t}}^+_{3\dot{5}}$ and $\hat{\boldsymbol{t}}^+_{4\dot{5}}$ need to vanish in order to find a consistent set of BPS equations in this case.  

As in all the previous cases, it turns out that the scalar potential vanishes. The superpotential is still given by \eqref{45SPot} for $\lambda=0$. This gives rise to the same set of BPS equations \eqref{45BPS1} to \eqref{45BPS6} with all additional scalar fields constant. The ansatz for the Killing spinors, in this case, takes the form of  
\begin{equation}
\epsilon_+=e^{\frac{A(r)}{2}+B_1(r)\gamma_{12}+B_2(r)\gamma_{34}}\epsilon_{+}^{0}\qquad\text{and}\qquad\epsilon_-=e^{\frac{A(r)}{2}-B_1(r)\gamma_{\dot{1}\dot{2}}-B_2(r)\gamma_{\dot{3}\dot{4}}}\epsilon_{-}^{0}.
\end{equation}
The BPS equations for $B_1$ and $B_2$ are given by
\begin{eqnarray}
B'_1&=&\frac{2ge^{-3\varphi-8\phi_0}\kappa\varsigma_1\sqrt{1+4\varsigma_2^2}}{\sqrt{1+4\varsigma_1^2}}\\ 
\textrm{and}\qquad
B'_2&=&\frac{2ge^{-3\varphi-8\phi_0}\kappa\varsigma_2\sqrt{1+4\varsigma_1^2}}{\sqrt{1+4\varsigma_2^2}}\, .
\end{eqnarray}
All these equations lead to the same solution given in \eqref{45DW1} to \eqref{45DW5} and \eqref{45DW7} for $\lambda=0$ together with
\begin{eqnarray}
B_1&=&C_6-\frac{1}{2}\tan^{-1}(2\varsigma_1)\\ 
\textrm{and}\qquad
B_2&=&C_7-\frac{1}{2}\tan^{-1}(2C_1\varsigma_1)
\end{eqnarray}
in which $C_6$ and $C_7$ are integration constants. For a particular case of $\varsigma_1=\varsigma_2=0$, we find the domain wall solution given in \eqref{45SpeDW} with $\lambda=0$ and the Killing spinors \eqref{DWKilling}. 

\section{Conclusions and discussions}\label{Discuss}
We have classified a number of gauge groups in six-dimensional maximal gauged supergravity with the embedding tensor in $\mathbf{5}^{+7}$, $\bar{\mathbf{5}}^{+3}$, $\mathbf{10}^{-1}$, $\mathbf{24}^{-5}$, and $\overline{\mathbf{45}}^{+3}$ representations of $GL(5)\subset SO(5,5)$. All these gauge groups are non-semisimple and consist mostly translational groups of various dimensions. We have also found supersymmetric domain wall solutions in $SO(2)\ltimes \mathbb{R}^8$, $SO(2)\ltimes \mathbb{R}^6$, $CSO(2,0,2)\ltimes \mathbb{R}^4$, $CSO(2,0,2)\ltimes \mathbb{R}^2$, and $CSO(2,0,1)\ltimes \mathbb{R}^4$ gauge groups. These solutions are either $\frac{1}{2}$- or $\frac{1}{4}$-supersymmetric preserving $SO(2)$ symmetry. Some of the resulting gauge groups could have higher dimensional origins in terms of Scherk-Schwarz reductions from seven dimensions or truncations of eleven-dimensional supergravity on twisted tori (possibly) with fluxes. The consistent gaugings identified in this paper to some extent enlarge the known gauge groups pointed out in \cite{6D_Max_Gauging} and those constructed in \cite{6D_DW_I}. These would hopefully be useful in the context of DW/QFT correspondence and related aspects. The domain wall solutions given here are exhaustive for all the gaugings under consideration at least for domain walls with any residual symmetry. These could also be added to the known classification of supersymmetric domain walls. It would be useful to find more general and more complicated solutions without any residual symmetry.   

Constructing truncation ansatze for embedding the domain wall solutions found here in string/M-theory using $SO(5,5)$ exceptional field theory given in \cite{SO5_5_EFT} is of particular interest. This would provide a complete framework for a holographic study of five-dimensional maximal SYM. It is also interesting to perform a similar analysis for gaugings under $SO(4,4)\subset SO(5,5)$ and construct the corresponding embedding tensors together with possible supersymmetric domain walls. These gaugings can be truncated to gaugings of half-maximal $N=(1,1)$ supergravity coupled to four vector multiplets in which supersymmetric $AdS_6$ vacua are known to exist \cite{D4D8,F4_flow,AdS6_Jan}. In this case, the results could be useful in the study of both DW$_6$/QFT$_5$ duality and AdS$_6$/CFT$_5$ correspondence as well. 

Finally, generalizing the standard domain wall solutions in this paper and \cite{6D_DW_I} to curved domain walls with non-vanishing vector and tensor fields could also be worth considering. This type of solutions describes conformal defects or holographic RG flows from five-dimensional $N=4$ super Yang-Mills theories to lower-dimensional (conformal) field theories via twisted compactifications. A number of similar solutions in seven dimensions have been given in \cite{7D_Defect, 7D_twist} while examples of holographic solutions dual to surface defects in five-dimensional $N=2$ SCFTs from $N=(1,1)$ gauged supergravity have appeared recently in \cite{6D_surface_defect}. 

\begin{acknowledgments}
This work is supported by the Second Century Fund (C2F), Chulalongkorn University. P. K. is supported by The Thailand Research Fund (TRF) under grant RSA6280022.
\end{acknowledgments}

\appendix
\section{Useful formulae}\label{AppA}
In this appendix, we collect some formulae and relations used in the main text, see \cite{6D_DW_I} for more detail. We begin with $SO(5,5)\rightarrow GL(5)$ branching rules used throughout the paper. With the branching rule for an $SO(5,5)$ vector representation
\begin{equation}\label{VecDec}
\mathbf{10}\ \rightarrow\ \mathbf{5}^{+2}\,\oplus\,\overline{\mathbf{5}}^{-2},
\end{equation}
we can decompose the $SO(5,5)$ generators as 
\begin{equation}
\boldsymbol{t}_{MN}\ \rightarrow\ (\boldsymbol{t}_{mn},{\boldsymbol{t}^m}_n,\boldsymbol{t}^{mn})
\end{equation}
with ${\boldsymbol{t}_m}^n=-{\boldsymbol{t}^n}_m$. 

In vector representation, $SO(5,5)$ generators $\boldsymbol{t}_{MN}=\boldsymbol{t}_{[MN]}$ can be chosen as
\begin{equation}
{(\boldsymbol{t}_{MN})_P}^Q\ =\ 4\eta_{P[M}\delta^Q_{N]}
\end{equation}
satisfying the Lie algebra 
\begin{equation}\label{SO(5,5)algebra}
\left[\boldsymbol{t}_{MN},\boldsymbol{t}_{PQ}\right]\ =\ 4(\eta_{M[P}\boldsymbol{t}_{Q]N}-\eta_{N[P}\boldsymbol{t}_{Q]M}).
\end{equation}
With the generators of the shift and hidden symmetries respectively denoted by $\boldsymbol{s}_{mn}=\boldsymbol{t}_{mn}$ and $\boldsymbol{h}^{mn}=\boldsymbol{t}^{mn}$, the $SO(5,5)$ generators can be written as
\begin{equation}
{(\boldsymbol{t}_{MN})_P}^Q=\begin{pmatrix} {\boldsymbol{t}^m}_n & \boldsymbol{h}^{mn} \\
								\boldsymbol{s}_{mn} & -{\boldsymbol{t}^n}_m \end{pmatrix}.
\end{equation}
In this form, we can see that the $GL(5)$ subgroup generated by ${\boldsymbol{t}_m}^n$ is embedded diagonally. 
\\
\indent The branching rules for spinor and conjugate spinor representations of $SO(5,5)$
\begin{equation}
\mathbf{16}_s\ \rightarrow\ \overline{\mathbf{5}}^{+3}\,\oplus\,\mathbf{10}^{-1}\,\oplus\,\mathbf{1}^{-5}\qquad\text{and}\qquad\mathbf{16}_c\ \rightarrow\ \mathbf{5}^{-3}\,\oplus\,\overline{\mathbf{10}}^{+1}\,\oplus\,\mathbf{1}^{+5}
\end{equation}
are realized respectively by the following relations
\begin{equation}
\Psi_A\ =\ \mathbb{T}_{Am}\Psi^m+\mathbb{T}_{A}^{mn}\Psi_{mn}+\mathbb{T}_{A\ast}\Psi_\ast
\end{equation}
with $\Psi_{mn}=\Psi_{[mn]}$ and
\begin{equation}
\Psi^A\ =\ \mathbb{T}^{Am}\Psi_m+\mathbb{T}^{A}_{mn}\Psi^{mn}+\mathbb{T}^{A}_{\ast}\Psi_\ast\, .
\end{equation}
The transformation matrices are defined by
\begin{eqnarray}
\mathbb{T}_{Am}&=&\frac{1}{2\sqrt{2}}(\Gamma_m)_{AB}\boldsymbol{p}^B_{\alpha\beta}\Omega^{\alpha\beta},\label{TranMatTDef1}\\
\mathbb{T}_{A}^{mn}&=&\frac{1}{4\sqrt{2}}{(\Gamma^{mn})_A}^B\boldsymbol{p}_B^{\alpha\beta}\Omega_{\alpha\beta},\label{TranMatTDef2}\\
\mathbb{T}_{A\ast}&=&\frac{1}{10}{({\Gamma^m}_m)_A}^B\boldsymbol{p}_B^{\alpha\beta}\Omega_{\alpha\beta}\label{TranMatTDef3}
\end{eqnarray}
with the inverse matrices of $\mathbb{T}_A$ are simply given by their complex conjugation $\mathbb{T}^A=(\mathbb{T}_A)^{-1}=(\mathbb{T}_A)^*$ satisfying the relations
\begin{eqnarray}
\mathbb{T}^{Am}\mathbb{T}_{An}\ =\ \delta^m_n,\qquad\quad\mathbb{T}^{A}_{mn}\mathbb{T}_{A}^{pq}& =& \delta^{[p}_{m}\delta^{q]}_{n},\qquad\mathbb{T}^{A}_\ast\mathbb{T}_{A\ast}\ =\ 1,\nonumber\\
\mathbb{T}^{Am}\mathbb{T}_{Anp}\ =\ 0,\qquad\quad\,\mathbb{T}^{Am}\mathbb{T}_{A\ast}& =& 0,\qquad\quad\mathbb{T}^{A}_{mn}\mathbb{T}_{A\ast}\ =\ 0
\end{eqnarray}
together with
\begin{equation}
\mathbb{T}^{Am}\mathbb{T}_{Bm}+\mathbb{T}^{A}_{mn}\mathbb{T}_{B}^{mn}+\mathbb{T}^{A}_\ast\mathbb{T}_{B\ast}=\delta^A_B\, .
\end{equation}
The transformation matrices $\boldsymbol{p}_A^{\alpha\beta}$ and inverse matrices $\boldsymbol{p}^A_{\alpha\beta}$ are given by
\begin{eqnarray}\label{thepmatrix}
\boldsymbol{p}_A^{\alpha\beta}&=&\delta_A^\alpha\delta_1^{\beta}+\delta_A^{\alpha+4}\delta_2^{\beta}+\delta_A^{\alpha+8}
\delta_3^{\beta}+\delta_A^{\alpha+12}\delta_4^{\beta},\nonumber\\
\boldsymbol{p}^A_{\alpha\beta}&=&\delta^A_\alpha\delta^1_{\beta}+\delta^A_{\alpha+4}\delta^2_{\beta}
+\delta^A_{\alpha+8}\delta^3_{\beta}+\delta^A_{\alpha+12}\delta^4_{\beta}\, .
\end{eqnarray}
These matrices satisfy the relations 
\begin{equation}
\boldsymbol{p}_A^{\alpha\beta}\boldsymbol{p}^B_{\alpha\beta}\ =\ \delta_A^B\qquad\text{and}\qquad\boldsymbol{p}_A^{\alpha\delta}\boldsymbol{p}^A_{\beta\gamma}\ =\ \delta^{\alpha}_{\beta}\delta^{\delta}_{\gamma}
\end{equation}
and can be used to express a spinor index of $SO(5,5)$, $A,B,\ldots$, in terms of a pair of $USp(4)$ fundamental or $SO(5)$ spinor indices $(\alpha\beta)$.

The matrices $(\Gamma_m)_{AB}$ appearing in the $SO(5,5)$ gamma matrices $\Gamma_M=\left(\Gamma_m,\Gamma^m\right)$ are related to $SO(5)\times SO(5)$ gamma matrices $\Gamma_{\ul{A}}=(\Gamma_a,\Gamma_{\dot{a}})$ as
\begin{equation}
(\Gamma_M)_{AB}={{\mathbb{M}}_M}^{\underline{A}}(\Gamma_{\underline{A}})_{AB}
\end{equation}
with
\begin{equation}
(\Gamma_{\underline{A}})_{AB}={(\Gamma_{\underline{A}})_{A}}^{C}\boldsymbol{c}_{CB}\, .
\end{equation}
The charge conjugation matrix $\boldsymbol{c}_{AB}$ is given by
\begin{equation}
\boldsymbol{c}_{AB}=\boldsymbol{p}_{A}^{\alpha\dot{\alpha}}\boldsymbol{p}_B^{\beta\dot{\beta}}
\Omega_{\alpha\beta}\Omega_{\dot{\alpha}\dot{\beta}}
\end{equation}
in which $\boldsymbol{p}_{A}^{\alpha\dot{\alpha}}$ and $\boldsymbol{p}^{A}_{\alpha\dot{\alpha}}$ are defined in parallel with \eqref{thepmatrix}. For convenience, we also note the transformation matrix $\mathbb{M}$ here
\begin{equation}\label{offDiagTrans}
\mathbb{M}\ =\ \frac{1}{\sqrt{2}}\begin{pmatrix} \mathds{1}_5 & \mathds{1}_5 \\ \mathds{1}_5 & -\mathds{1}_5 \end{pmatrix}.
\end{equation}
\indent The $SO(5)\times SO(5)$ gamma matrices can be written as 
\begin{equation}
{(\Gamma_{\underline{A}})_{A}}^{B}=\left({(\gamma_a)_{A}}^{B},{(\gamma_{\dot{a}})_{A}}^{B}\right)
\end{equation}
in which ${(\gamma_a)_{A}}^{B}$ and ${(\gamma_{\dot{a}})_{A}}^{B}$ are explicitly given by
\begin{eqnarray}
{(\gamma_a)_{A}}^{B}&=&\boldsymbol{p}_A^{\alpha\dot{\alpha}}{(\gamma_a)_{\alpha\dot{\alpha}}}^{\beta\dot{\beta}}\boldsymbol{p}^B_{\beta\dot{\beta}}\ = \ \boldsymbol{p}_A^{\alpha\dot{\alpha}}{(\gamma_a)_\alpha}^\beta\delta_{\dot{\alpha}}^{\dot{\beta}}\boldsymbol{p}^B_{\beta\dot{\beta}},\\
{(\gamma_{\dot{a}})_{A}}^{B}&=&\boldsymbol{p}_A^{\alpha\dot{\alpha}}{(\gamma_{\dot{a}})_{\alpha\dot{\alpha}}}^{\beta\dot{\beta}}\boldsymbol{p}^B_{\beta\dot{\beta}}\ = \ \boldsymbol{p}_A^{\alpha\dot{\alpha}}\delta_\alpha^\beta{(\gamma_{\dot{a}})_{\dot{\alpha}}}^{\dot{\beta}}
\boldsymbol{p}^B_{\beta\dot{\beta}}\, .
\end{eqnarray}
With all these, the matrices $(\Gamma_m)_{AB}$ can be explicitly defined in terms of the $SO(5)$ gamma matrices as
\begin{equation}\label{GammEx}
(\Gamma_m)_{AB}=\frac{1}{\sqrt{2}}\boldsymbol{p}_A^{\alpha\dot{\alpha}}\left[{(\gamma_m)_\alpha}^\beta\delta_{\dot{\alpha}}^{\dot{\beta}}
+\delta_\alpha^\beta{(\gamma_m)_{\dot{\alpha}}}^{\dot{\beta}}\right]\Omega_{\beta\gamma}\Omega_{\dot{\beta}\dot{\gamma}}
\boldsymbol{p}_B^{\gamma\dot{\gamma}}\, .
\end{equation}
\indent Finally, ${(\Gamma^{mn})_A}^B$ and ${({\Gamma^m}_m)_A}^B$ matrices are given by
\begin{equation}
{(\Gamma^{mn})_A}^B=\frac{1}{2}\left[{(\Gamma^m)_A}^{C}{(\Gamma^n)_{C}}^B-{(\Gamma^n)_A}^{C}{(\Gamma^m)_{C}}^B\right]
\end{equation}
and
\begin{equation}\label{Gammamm}
{({\Gamma^m}_m)_A}^B=\frac{1}{2}\left[{(\Gamma^m)_A}^{C}{(\Gamma_m)_{C}}^B-{(\Gamma_m)_A}^{C}{(\Gamma^m)_{C}}^B\right]
\end{equation}
with
\begin{eqnarray}\label{GammUpDown}
{(\Gamma_m)_A}^B&=&\frac{1}{\sqrt{2}}\boldsymbol{p}_A^{\alpha\dot{\alpha}}\left[{(\gamma_m)_\alpha}^\beta\delta_{\dot{\alpha}}^{\dot{\beta}}
+\delta_\alpha^\beta{(\gamma_m)_{\dot{\alpha}}}^{\dot{\beta}}\right]\boldsymbol{p}^B_{\beta\dot{\beta}},\\
{(\Gamma^m)_A}^B&=&\frac{1}{\sqrt{2}}\boldsymbol{p}_A^{\alpha\dot{\alpha}}\left[{(\gamma_m)_\alpha}^\beta\delta_{\dot{\alpha}}^{\dot{\beta}}
-\delta_\alpha^\beta{(\gamma_m)_{\dot{\alpha}}}^{\dot{\beta}}\right]\boldsymbol{p}^B_{\beta\dot{\beta}}\, .
\end{eqnarray}
In this paper, we use the following representation for $SO(5)$ gamma matrices
\begin{eqnarray}
\gamma_1&=&-\sigma_2\otimes\sigma_2,\quad\, \gamma_2\ =\ \mathds{1}_2\otimes\sigma_1,\quad
\gamma_3\ =\ \mathds{1}_2\otimes\sigma_3,\nonumber\\ 
\gamma_4&=&\sigma_1\otimes\sigma_2,\qquad \gamma_5\ =\ \sigma_3\otimes\sigma_2
\end{eqnarray}
in which $\sigma_1$, $\sigma_2$ and $\sigma_3$ are the usual Pauli matrices given by
\begin{equation}
\sigma_1\ = \ \begin{pmatrix} 0 & 1\\ 1 & 0 \end{pmatrix},\qquad \sigma_2\ = \ \begin{pmatrix} 0 & -i\\ i & 0 \end{pmatrix},\qquad\sigma_3\ = \ \begin{pmatrix} 1 & 0\\ 0 & -1 \end{pmatrix}\, .
\end{equation}


\end{document}